\documentclass[journal,twocolumn,10pt]{IEEEtran}
\IEEEoverridecommandlockouts 
\usepackage{epsfig,latexsym}
\usepackage{float}
\usepackage{indentfirst}
\usepackage{amsmath}
\usepackage{amssymb}
\usepackage{times}
\usepackage{subfigure}
\usepackage{pifont}
\usepackage{psfrag}
\usepackage{cite}
\usepackage{url}
\usepackage{lastpage}
\usepackage{bm}
\usepackage{color}
\usepackage{fancyhdr}
\usepackage[center]{caption2}
\usepackage{balance}
\usepackage{fancyhdr,lastpage}
\usepackage{hyperref}
\begin{document}
\title{Simultaneous Wireless Information Power Transfer for MISO Secrecy Channel}
\author{\IEEEauthorblockN{Zheng Chu, Martin Johnston, and St\'{e}phane Le Goff}	 \\
\vspace{-0.3in}
\thanks{
The authors are with the School of Electrical and Electronic Engineering, Newcastle University, Newcastle upon Tyne, NE1
7RU, U.K. (Emails: z.chu@ncl.ac.uk; martin.johnston@ncl.ac.uk; stephane.le-goff@ncl.ac.uk)} }
\vspace{-0.2in}
\clearpage\maketitle
\thispagestyle{empty}
\begin{abstract}
This paper investigates simultaneous wireless information and power transfer (SWIPT) for multiuser multiple-input-single-output (MISO) secrecy channel. First, transmit beamfoming without artificial noise (AN) design is considered, where two secrecy rate optimization frameworks (i.e., secrecy rate maximization and harvested energy maximization) are investigated. These two optimization problems are not convex, and cannot be solved directly. For secrecy rate maximization problem, we employ bisection method to optimize the associated power minimization problem, and first-order Taylor series expansion is consider to approximate the energy harvesting (EH) constraint and the harvested energy maximization problem. Moreover, we extend our proposed algorithm to the associated robust schemes by incorporating with channel uncertainties, where two-level method is proposed for the harvested energy maximization problem. 
Then, transmit beamforming with AN design is studied for the same secrecy rate maximization problem, which are reformulated into semidefinite programming (SDP) based on one-dimensional search and successive convex approximation (SCA), respectively. Moreover, tightness analysis of rank relaxation is provided to show the optimal transmit covariance matrix exactly returns rank-one. Simulation results is provided to validate the performance of the proposed algorithm.    
\end{abstract}
\IEEEpeerreviewmaketitle
\setlength{\baselineskip}{1\baselineskip}
\newtheorem{definition}{Definition}
\newtheorem{fact}{Fact}
\newtheorem{assumption}{Assumption}
\newtheorem{theorem}{Theorem}
\newtheorem{lemma}{Lemma}
\newtheorem{corollary}{Corollary}
\newtheorem{proposition}{Proposition}
\newtheorem{example}{Example}
\newtheorem{remark}{Remark}
\newtheorem{algorithm}{Algorithm}
\newcommand{\mv}[1]{\mbox{\boldmath{$ #1 $}}}
\section{Introduction}\label{section introduction}
As one of the main techniques in fifth-generation (5G)
wireless networks, energy harvesting is a technique to extract energy from the external natural environment (i.e., solar power, wind energy, etc.), which can be applied to circumvent energy-limited issue and improve the energy efficiency of wireless networks\cite{Hossain_5G_WCM,Emergy_Tech_M06}. However, traditional energy harvesting techniques are dependent on external sources, which are not part of the communication network. Since it is not always possible that the nature environment
can provide a stable energy to the wireless devices, the energy can be alternatively provided by wireless devices, since radio frequency (RF) signals can carry energy that is used as a vehicle for transporting information \cite{Zhang_Rui_SWIPT_TWC13}. In particular,
the transmitter in a communication network not only can
transmit the signal, but also transfer the power to the receivers who are employed to harvest energy to charge for their devices. Wireless power transfer has been a promising paradigm to provide power supplies for communication devices to mitigate the energy scarcity and extend the lifetime of wireless networks \cite{Varshney_08,Shannon_Tesla_10}. \\
Secrecy performance played more and more important role in wireless networks, and there is an increasing number of interests focusing on physical-layer security. Unlike traditional cryptographic methods, which is employed to improve secrecy performance in the network layer, physical-layer security was developed in terms of information-theoretical aspects improve the security of wireless transmission \cite{Liang_J08,Poor_Sig_Process_J10,Poor_Info_Theory_J11}.  Multiple antenna wiretap channel has been widely considered with the advantage of having extra degrees of freedom and diversity gains \cite{Wornell_Info_Theory1_J10}. In addition, some state-of-art techniques have been developed for multiple-antenna transceivers, which aims to introduce more interference to eavesdroppers, including cooperative beamforming (CB), cooperative jammer (CJ), artificial noise (AN), energy harvesting beamforming (EHB), device-to-device transmission (D2D), etc \cite{Poor_Sig_Process_J10,Ma_Sig_Process_J11,Kim_CB_Sec_J11,Kim_CB_AF_Sec_J12,KK_Wong_Sig_Process_J11,Huang_Jing_TSP_CJ_11,
Zheng_Secrecy_J15,Ma_TSP_J13,Zhang_Rui_TSP_Sec_SWIPT_J14,Khandaker_TIFS_J15,chu2014robust}.  
The convex optimization techniques have been employed to solve the secrecy rate optimization problem (i.e., power minimization and secrecy rate maximization) based on perfect and imperfect channel state information (CSI) \cite{Ma_Sig_Process_J11}. The cooperative beamforming is employed that requires the relays to forward the signal from the source to the legitimate user based on the assumption that the direct transmission is not available. The optimal power allocation for single-relay multicarrier case based on decode-and-forward (DF) has been proposed to maximize the sum secrecy rate \cite{Kim_CB_Sec_J11}, whereas in \cite{Kim_CB_AF_Sec_J12} the relay works in amplify-and-forward (AF) scheme for MIMO system, where the source and relay beamforming is jointly designed to maximize the secrecy rate in the cooperative scheme. Cooperative jamming is applied in improving physical layer security  \cite{KK_Wong_Sig_Process_J11,Zheng_Secrecy_J15,Huang_Jing_TSP_CJ_11}. For single-antenna case, the secrecy rate is maximized by employing one-dimensional search algorithm \cite{KK_Wong_Sig_Process_J11}. In \cite{Zheng_Secrecy_J15}, Taylor series expansion is investigated to approximate the secrecy rate optimization for MIMO channel with cooperative jammer, which can be reformulate the non-convex secrecy rate optimization framework to convex form, in addition, game theory technique (i.e., \emph{Stackelberg} game) is applied for improving secrecy rate, and \emph{Stackelberg} game equilibrium can be derived in terms of closed-form solutions.
For MIMO relay networks, the optimal power
allocation is derived for the generalized singular value decomposition (GSVD)-based secure relaying schemes\cite{Huang_Jing_TSP_CJ_11}. Joint optimization of transmit beamforming with artificial noise (AN) is proposed in \cite{Ma_TSP_J13}, whereas the energy harvesting beamforming (EHB) and D2D transmission are not only improve the secrecy rate, but also can harvest the energy to prolong the lifetime of the devices by recharging the batteries and enhancing D2D transmission rate, respectively \cite{Zhang_Rui_TSP_Sec_SWIPT_J14,Khandaker_TIFS_J15,chu2014robust}.   
\\
More and more recent attentions are focusing on the combination of transmit beamforing with AN, where AN can be employed to mask the transmit beamforming so that improve the secrecy rate, in addition, the transmit beamforming and AN can be harvested by the energy harvesting (EH) receivers to improve energy harvesting performance \cite{Zhang_Rui_TSP_Sec_SWIPT_J14,Li_Qiang_ICCASP2014_C14,Khandaker_TIFS_J15}. Moreover, the eavesdroppers in these two literatures play the two roles: a) passive eavesdropping, b) energy harvesting. The secrecy rate maximization problem are formulated into semidefinite programming (SDP) and provide the proof of the rank-one for the transmit covariance matrix. Robust secure transmission for MISO Simultaneous Wireless Information and Power Transfer (SWIPT) system are proposed without AN \cite{Feng_Renhai_TVT_J15} and with AN \cite{Tian_Maoxin_SPL_J15}, respectively by incorporating with channel uncertainties, where the proofs of the rank-two solution for the transmit covariance matrix have been provided. \\
Motivated by the existing literatures \cite{Feng_Renhai_TVT_J15,Tian_Maoxin_SPL_J15}, this paper considers SWIPT for MISO secrecy system with multiple eavesdroppers and multiple EH receivers, here the secure transmit beamforming without AN and with AN are designed. The contribution of this paper is that we address the following problems:
\begin{enumerate}
\item \emph{Secrecy rate optimization without AN:}
We formulate the secrecy rate optimization framework to design the transmit beamforming without AN. Two optimization problems (i.e., secrecy rate maximization and harvested energy maximization) are consider, which are not convex and cannot be solved directly. In order to circumvent this non-convexity issue, we propose a novel reformulation for the secrecy rate maximization problem based on bisection method, and the harvested energy maximization problem via first-order Taylor series based one-dimensional line search algorithm, respectively. In addition, these associated robust schemes are solved by incorporating with channel uncertainties.
Unlike the most existing literature, which employs semidefinite relaxation (SDR) (or rank-relaxation) to design the transmit beamforming, in this paper, we consider the relaxation to circumvent the rank-relaxation and we show our proposed schemes can achieve the same secrecy performance with that via SDR.  
\item \emph{Secrecy rate optimization with AN:}
Secrecy rate maximization problem is formulated to jointly optimization for transmit beamforming with AN. In this paper, we propose two reformulations for this optimization framework based on one-dimensional line search and SCA algorithms, respectively, which can be used to reformulate the problem to semidefinite programming (SDP). In addition, we provide the tightness analysis for the rank-relaxation, which shows the optimal solution of the secrecy rate maximization exactly returns rank-one. Moreover, we extend our proposed schemes to the associated case by incorporating with channel uncertainties and the case with multiantenna eavesdroppers and EH receivers.    
\end{enumerate}
The remaining part of this paper is organized as follows. A system model is described in Section \ref{section system model}. The secrecy rate optimization problems with SWIPT are solved to jointly optimize the transmit beaforming without AN based on perfect and imperfect CSI in Section \ref{section Secrecy_Rate_Optimization_without_AN} and \ref{section Robust_Secrecy_Rate_Optimization_with_Channel_Uncertainty}. 
Secrecy rate optimization problem with SWIPT to joint optimization of the transmit beaforming with AN is investigated in Section \ref{section Secure_SWIPT_of_Joint_Beamforming_and_Artificial_Noise_Design}.
Section \ref{section Numerical_results} provides the simulation results to support the proposed schemes. Finally, conclusions are presented in Section \ref{section Conclusions}.
\subsection{Notation}
\indent We use the upper case boldface letters for matrices and lower case boldface letters for vectors. $(\cdot)^{T}$, $(\cdot)^{*}$ and $(\cdot)^{H}$ denote the transpose, conjugate and conjugate transpose respectively. Tr$(\cdot)$ and $\mathbb{E}\{\cdot\}$ stand for trace of a matrix and the statistical expectation for random variables. Vec$(\mathbf{A})$ is the vector obtained by stacking the columns of $\mathbf{A}$ on top of one another and $\otimes$ is the Kronecker product. $\mathbf{A}\succeq 0$ indicates that $\mathbf{A}$ is a positive semidefinite matrix. $\mathbf{I}$ and $(\cdot)^{-1}$ denote the identity matrix  with appropriate size and the inverse of a matrix respectively. $\|\cdot\|_{2}$ represents the Euclidean norm of a matrix. $\Re\{\cdot\}$ stands for the real part of a complex number, whereas $|\mathbf{A}|$ denotes the determinant of $\mathbf{A}$. $[x]^{+}$ represents $\max\{x,0\}$. 
\section{System Model}\label{section system model}
We consider a MISO secured SWIPT channel, where it consists of one multiantenna legitimate transmitter, one legitimate user, $ K $ eavesdroppers and $ L $ energy harvesting (EH) receiver. It is assumed that the transmitter is equipped with $ N_{T} $ transmit antennas, whereas the legitimate user, the eavesdroppers and the EH receivers each have a single receive antenna. The channel coefficients between the legitimate transmitter and the legitimate user, the $ k $-th eavesdropper as well as the $ l $-th EH receiver are denoted by $ \mathbf{h}_{s} \in \mathbb{C}^{N_{T} \times 1} $, $ \mathbf{h}_{e,k} \in \mathbb{C}^{N_{T} \times 1} $, and $ \mathbf{h}_{p,l} \in \mathbb{C}^{N_{T} \times 1} $, respectively. The noise power at the legitimate user and the eavesdroppers are assumed to be $ \sigma_{s}^{2} $ and $ \sigma_{e}^{2} $. The received signal at the legitimate user and the $ k $-th eavesdropper can be written as 
{\small \begin{eqnarray}
y_{s} &\!\!\!\!\!=&\!\!\!\!\! \mathbf{h}_{s}^{H}\mathbf{w}\mathbf{s} \!+\! \mathbf{n}_{s},\nonumber\\
y_{e,k} &\!\!\!\!\!=&\!\!\!\!\! \mathbf{h}_{e,k}^{H}\mathbf{w}\mathbf{s} \!+\! \mathbf{n}_{e,k},~ k = 1,...,K, \nonumber
\end{eqnarray} }
where $ s (\mathbb{E}\{ s^{2} \} = 1) $ and $ \mathbf{w}  \in \mathbb{C}^{N_{T} \times 1} $ are the desired signal for the legitimate user and the beamforming at the legitimate transmitter, respectively. In addition, $ \mathbf{n}_{s} \sim \mathcal{CN}(0, \sigma_{s}^{2}) $ and $ \mathbf{n}_{e,k} \sim \mathcal{CN}(0, \sigma_{e}^{2}) $ represent the noise of the legitimate user and the $ k $-th eavesdropper, respectively. Thus, the achieved secrecy rate at the legitimate user is expressed as follows:
{\small \begin{eqnarray}
\!\!\!\!\!\!R_{s} \!=\! \left[\log\bigg( 1 \!+\! \frac{|\mathbf{h}_{s}^{H}\mathbf{w}|^{2}}{\sigma_{s}^{2}} \bigg) - \max_{k} \log\bigg( 1 \!+\! \frac{|\mathbf{h}_{e,k}^{H}\mathbf{w}|^{2}}{\sigma_{e}^{2}} \bigg)\right]^{+}\!\!,~ \forall k.\!\!\!\!\!\!\!\!\!\!\!\!\!\!\!\!\!\!\!\nonumber\\
\end{eqnarray} }
The harvested energy at the $ l $-the EH receiver is written as
{\small \begin{eqnarray}
E_{l} = |\mathbf{h}_{l}^{H}\mathbf{w}|^{2},~ \forall l.
\end{eqnarray} }
\emph{Remark:} This system model is considered to consists of $ L $ EH receivers, which harvest a number amount of power carried by RF signal without AN or with AN based on a reliable transmission scenario. These EH receivers sometimes play a `` helper '' role by employing the harvested power to introduce a jamming signal to confuse the eavesdroppers \cite{Xing_H_GLOBECOM_C2014}. However, the efficiency of this harvest-and-jamming policy is dependant on the network topology \cite{Zhiguo_Ding_CM_J14}. In this paper, we mainly focus on the secrecy rate optimization without AN or with AN, based on this reliable communication, the performance the EH receiver harvest the power is exploited, whereas how to use the harvested power is beyond our paper.

\section{Secrecy Rate Optimization without AN}\label{section Secrecy_Rate_Optimization_without_AN}
In this section, we consider the MISO secrecy with multiple eavesdroppers and EH receivers shown in Section \ref{section system model}, where the secure beamforming is designed without AN, in addition, the secrecy rate optimization frameworks (i.e., secrecy rate maximization and harvested energy maximization) are reformulated as a novel SDP form to circumvent rank relaxation. 
Thus, these optimization problems are formulated as follows:
\begin{itemize}
\item Secrecy rate maximization: The secrecy rate is maximized subject to transmit power and minimum harvested energy constraints.
{\small \begin{eqnarray}\label{eq:sec_rate_max_ori}
\max_{\mathbf{w}} \min_{k} R_{s}, ~
s.t. ~\|\mathbf{w}\|^{2} \leq P,~
\min_{l} E_{l} \geq E, ~\forall k, l,
\end{eqnarray} }
where $ P $ is the maximum available transmit power at the legitimate transmitter, $ E $ denotes the target harvested energy of the EH receivers.   
\item Harvested energy maximization: The harvested energy is maximized with the constraints of the secrecy rate and transmit power.
{\small \begin{eqnarray}\label{eq:harvested_energy_max_ori}
\max_{\mathbf{w}} \min_{l} E_{l}, ~
s.t. ~ \min_{k} R_{s,k} \geq R,~\|\mathbf{w}\|^{2} \leq P, ~\forall k, l,
\end{eqnarray} }
where $ R $ is the predefined secrecy rate of the legitimate user.
\end{itemize}
These above problems are not convex, and cannot be solved directly. Unlike existing literature \cite{Feng_Renhai_TVT_J15}, where rank relaxation is consider to reformulate the secrecy rate, however, since it is challenging to show rank-one solution, the authors provided suboptimal solution for the secrecy rate maximization. In this paper, we propose a novel relax method for secrecy rate constraint to circumvent rank relaxation.
\subsection{Secrecy Rate Maximization}\label{section Secrecy_Rate_Maximization}
In this subsection, we consider the relax of the secrecy rate maximization problem \eqref{eq:sec_rate_max_ori}, which is not convex due to the secrecy rate and non-convex EH constraint, and cannot be solved directly. 
In order to circumvent these two issues, we convert this problem into a sequence of the power minimization problems, one for each target rate $ R > 0 $. The optimal solution of the secrecy rate maximization problem can be obtained by solving the corresponding power minimization problem with different $ R $, which can be obtained by using bisection search over $ R $. 
Thus, the bisection method can be summarized in Table \ref{Table:bisection_method} to solve this secrecy rate maximization problem.
\begin{table}
\vspace{-0.05in}
\hrule
\caption{\label{Table:bisection_method} Bisection methods}
\vspace{0.05in}
\hrule 
\vspace{0.01in}
\begin{enumerate}
   \item Given lower and upper bound of the targeted secrecy rate $R_{\min}$ and $R_{\max}$, and a desired solution accuracy $ \tau $ (very small value).
   \item Setting $ R = (R_{\min} + R_{\max})/2 $.
  \item \textbf{Iteration loop begin}
       \begin{enumerate}
        \vspace{-1mm}     
        \item Solve the corresponding power minimization problem in \eqref{eq:power_min_ori} using the relaxation method to obtain the beamforming $ \mathbf{w} $.
        \item Computing the transmit power $ \tilde{P} = \|\mathbf{w}\|^{2}$.
        \item If $ \tilde{P} \leq P $, then $ R_{\min} = R $; otherwise, $ R_{\max} = R $. 
        \item \textbf{Until} $ R_{\max} - R_{\min} \leq \tau $, \textbf{break}.
      \end{enumerate}
        \item \textbf{Iteration loop end}
        \item $ R $ is the achieved secrecy rate of the secrecy rate maximization problem, and $ \mathbf{w} $ is the corresponding optimal solution.
\end{enumerate}
\hrule
\vspace{0.1in}
\end{table}
In the following, we will focus on the power minimization problem written as follows:
{\small \begin{eqnarray}\label{eq:power_min_ori}
\min_{\mathbf{w}} &\!\!\!\!&\!\!\!\! \|\mathbf{w}\|^{2}, \nonumber\\
s.t.  &\!\!\!\!&\!\!\!\!  \min_{k} R_{s} \geq R, ~\min_{l} E_{l} \geq E, ~\forall k,l.
\end{eqnarray} }
Now, we consider the power minimization problem based on the assumption that the transmitter has perfect CSI of the legitimate user, the eavesdroppers and the EH receivers. Thus, the problem in \eqref{eq:power_min_ori} can be relaxed as 
{\small \begin{eqnarray}\label{eq:power_min_modified}
\min_{\mathbf{w}} &\!\!\!\!&\!\!\!\! \|\mathbf{w}\|^{2}, \nonumber\\ 
s.t. &\!\!\!\!&\!\!\!\! \log\bigg( 1 \!+\! \frac{|\mathbf{h}_{s}^{H}\mathbf{w}|^{2}}{\sigma_{s}^{2}} \bigg) \!-\! \log\bigg( 1 \!+\! \frac{|\mathbf{h}_{e,k}^{H}\mathbf{w}|^{2}}{\sigma_{e}^{2}} \bigg) \!\geq\! R, ~ \forall k, \nonumber\\
&\!\!\!\!&\!\!\!\! |\mathbf{h}_{l}^{H}\mathbf{w}|^{2} \!\geq\! E, ~\forall l. 
\end{eqnarray} }
The above problem is not convex in terms of $ \mathbf{w} $ and the non-convex EH constraint. In order to circumvent these issues, we convert the secrecy rate constraint into SDP form and reformulate the EH constraint based on successive convex approximation (SCA), respectively. Thus, the following lemma is required to tackle the power minimization problem \eqref{eq:power_min_modified}:\\\\
\begin{lemma}\label{lemma:Schur_complement_applications}
The problem in \eqref{eq:power_min_modified} is reformulated into the following form:
{\small \begin{eqnarray}\label{eq:power_min_SCA}
\min_{\mathbf{w}} &\!\!\!\!&\!\!\!\! s_{1}, \nonumber\\ 
s.t. &\!\!\!\!&\!\!\!\! \left[\begin{array}{cc}
 s_{1} \\ \mathbf{w}  
\end{array} 
 \right] \succeq_{K} \mathbf{0}, \nonumber\\ 
 &\!\!\!\!&\!\!\!\! \mathbf{S}_{k} \!=\! \left[\!\!\begin{array}{cc}
 \frac{1}{\sigma_{s}}\mathbf{w}^{H}\mathbf{h}_{s}\mathbf{I} \!\!&\!\! \left[\begin{array}{cc}
\frac{2^{\frac{R}{2}}}{\sigma_{e}}\mathbf{w}^{H}\mathbf{h}_{e,k} \\ (2^{R} - 1)^{\frac{1}{2}}
 \end{array}
\right]   \\ 
 \left[\begin{array}{cc}
\frac{2^{\frac{R}{2}}}{\sigma_{e}}\mathbf{w}^{H}\mathbf{h}_{e,k} \\ (2^{R} - 1)^{\frac{1}{2}}
 \end{array}
\right]^{H} \!\!&\!\! \frac{1}{\sigma_{s}}\mathbf{w}^{H}\mathbf{h}_{s}
\end{array} 
 \!\!\right] \succeq \mathbf{0}, ~\forall k, \nonumber\\ 
&\!\!\!\!&\!\!\!\! x_{l} = \Re\{ \mathbf{w}^{H}\mathbf{h}_{l} \}, ~y_{l} = \Im\{ \mathbf{w}^{H}\mathbf{h}_{l} \}, ~\mathbf{u}_{l} = [x_{l} ~ y_{l}], \nonumber\\ 
&\!\!\!\!&\!\!\!\! \|\mathbf{u}_{l}^{(n)}\|^{2} 
+ 2 \sum_{i = 1}^{2} \mathbf{u}_{l}^{(n)}(i) [ \mathbf{u}_{l}(i) - \mathbf{u}_{l}^{(n)}(i) ] \geq E, ~\forall l.  
\end{eqnarray} }
\end{lemma}
\begin{IEEEproof}
Please refer to Appendix \ref{appendix:Schur_complement_applications}.
\end{IEEEproof}
In problem \eqref{eq:power_min_SCA}, the secrecy rate constraint is reformulated into SDP form, whereas the EH constraint is approximated by first-order Taylor series expansion. An initialization value of the vector $ \mathbf{u}_{l} $ is given by random generation, and $ \mathbf{u}_{l} $ can be updated at each iteration, and $ \mathbf{u}_{l}^{(n+1)} = \mathbf{u}_{l}^{(n)} $ holds when the algorithm converges, in addition, it is guaranteed
to converge to a locally optimal solution (quite close to the globally optimal solution) \cite{Hanif_SPL_J12,Hanif_SPL_J14}.
\vspace{-2em}
\subsection{Harvested Energy Maximization}
Now, we turn our attention to harvested energy maximization problem \eqref{eq:harvested_energy_max_ori}, which can be relaxed as 
{\small \begin{subequations}\label{eq:Energy_max_with_slack_variable}
\begin{eqnarray}
\max_{\mathbf{w}, t} &\!\!\!\!&\!\!\!\! t \nonumber\\ 
s.t. &\!\!\!\!&\!\!\!\! \|\mathbf{w}\|^{2} \leq P, \nonumber\\
&\!\!\!\!&\!\!\!\! \min_{l} |\mathbf{w}^{H}\mathbf{h}_{l}|^{2} \geq t,~ \forall l, \label{eq:Energy_max_energy_constraint}\\
&\!\!\!\!&\!\!\!\! \mathbf{S}_{k} 
 \succeq \mathbf{0}, ~\forall k, \label{eq:Energy_max_sec_rate_constraint}
\end{eqnarray}
\end{subequations} }
where $ t $ is a slack variable for the minimum EH constraint, $ \mathbf{S}_{k} $ has been defined in \eqref{eq:power_min_SCA}.
Although the secrecy rate constraint \eqref{eq:Energy_max_sec_rate_constraint} is LMI, the above problem is still not convex in terms of the harvested energy constraint \eqref{eq:Energy_max_energy_constraint}. 
The approximation is similar to \emph{Lemma 1} in Section \ref{section Secrecy_Rate_Maximization}, in which the energy maximization problem can be reformulated as 
{\small \begin{subequations}\label{eq:Energy_max_SCA01}
\begin{eqnarray}
\max_{\mathbf{w}, t} &\!\!\!\!&\!\!\!\! t \nonumber\\ 
s.t. &\!\!\!\!&\!\!\!\! \|\mathbf{w}\|^{2} \leq P, \nonumber\\
&\!\!\!\!&\!\!\!\! \mathbf{S}_{k} 
 \succeq \mathbf{0}, ~\forall k, \label{eq:Energy_max_sec_rate_constraint} \\
 &\!\!\!\!&\!\!\!\!  \|\mathbf{u}_{l}^{(n)}\|^{2} + 2\sum_{i=1}^{2} \mathbf{u}_{l}^{(n)}(i) [\mathbf{u}_{l}(i) - \mathbf{u}_{l}^{(n)}(i)] \geq t, ~\forall l, \label{eq:Energy_max_energy_constraint}
\end{eqnarray}
\end{subequations} }
where $ \mathbf{u}_{l} $ has been defined in \emph{Lemma} \ref{lemma:Schur_complement_applications}. 
The above problem is convex for $ \mathbf{u}^{(n)}_{l} $ at each iteration, and can be solved by interior-point method. 
\section{Robust Secrecy Rate Optimization with Channel Uncertainty} \label{section Robust_Secrecy_Rate_Optimization_with_Channel_Uncertainty}
In the previous section, we have solve the secrecy rate optimization problem based on the assumption that the transmitter has perfect CSI of the eavesdroppers. However, it is not always possible to have perfect CSI due to lack of cooperation as well as channel estimation and quantization error. Thus, we consider robust scheme based on the worst-case secrecy rate by incorporating channel uncertainties.
\subsection{Channel Uncertainty}\label{subsection channel uncertainty}
In this subsection, it is assumed that the channel state information is not available at the legitimate transmitter. The channel uncertainties based on worst-case scheme is modelled as  
{\small \begin{eqnarray}
\mathbf{h}_{s} &\!\!\!\!=&\!\!\!\! \mathbf{\bar{h}}_{s} \!+\! \mathbf{e}_{s},\nonumber\\ 
\mathbf{h}_{e,k} &\!\!\!\!=&\!\!\!\! \mathbf{\bar{h}}_{e,k} \!+\! \mathbf{e}_{e,k},\nonumber\\ 
\mathbf{h}_{l} &\!\!\!\!=&\!\!\!\! \mathbf{\bar{h}}_{l} \!+\! \mathbf{e}_{l}, \nonumber
\end{eqnarray} }
where $ \mathbf{\bar{h}}_{s} $, $ \mathbf{\bar{h}}_{e,k} $ and $ \mathbf{\bar{h}}_{l} $ denote the estimated channels of the legitimate user, the $ k $-th eavesdropper and the $ l $-th EH receiver, and $ \mathbf{e}_{s} $, $ \mathbf{e}_{e,k} $ and $ \mathbf{e}_{l} $ represent the corresponding channel errors, which are assumed to be bounded as 
{\small \begin{eqnarray}
\|\mathbf{e}_{s}\|_{2} &\!\!\!\!\!=&\!\!\!\!\! \|\mathbf{h}_{s} \!-\! \mathbf{\bar{h}}_{s}\|_{2} \!\leq\! \varepsilon_{s}, \textrm{for}~~ \varepsilon_{k} \geq 0,\nonumber\\
\|\mathbf{e}_{e,k}\|_{2} &\!\!\!\!\!=&\!\!\!\!\! \|\mathbf{h}_{e,k} \!-\! \mathbf{\bar{h}}_{e,k}\|_{2} \!\leq\! \varepsilon_{k}, \textrm{for}~~ \varepsilon_{k} \geq 0,\nonumber\\
\|\mathbf{e}_{l}\|_{2} &\!\!\!\!\!=&\!\!\!\!\! \|\mathbf{h}_{l} \!-\! \mathbf{\bar{h}}_{l}\|_{2} \!\leq\! \varepsilon_{l}, \textrm{for}~~ \varepsilon_{l} \geq 0. \nonumber
\end{eqnarray} }
In the following subsections, we will reformulate the secrecy rate optimization frameworks based on this channel uncertainties.
\subsection{Robust Secrecy Rate Optimization with Channel Uncertainty}
 Now, we solve the robust power minimization problem by incorporating with channel uncertainties, where the CSI of the legitimate user, the eavesdroppers and the EH receivers are not available at the legitimate transmitter. Thus, the relaxed robust problem is formulated as follows:
{\small \begin{subequations}\label{eq:robust_power_min_ori_full_channel_uncertainty}
\begin{eqnarray}
\min_{\mathbf{w}} &\!\!\!\!&\!\!\!\! \|\mathbf{w}\|^{2}, \nonumber\\
s.t. &&\!\!\!\!\!\!\!\!\!\! \min_{\mathbf{e}_{s}}\log\bigg( 1 \!+\! \frac{|(\mathbf{\bar{h}}_{s}+\mathbf{e}_{s})^{H}\mathbf{w}|^{2}}{\sigma_{s}^{2}} \bigg) \!-\! \nonumber\\
&&\!\!\!\!\!\!\!\! \max_{\mathbf{e}_{e,k}}\log\bigg( 1 \!+\! \frac{|(\mathbf{\bar{h}}_{e,k}+\mathbf{e}_{e,k})^{H}\mathbf{w}|^{2}}{\sigma_{e}^{2}} \bigg) \!\geq\! R, ~ \forall k, \label{eq:sec_rate_constraint_full}\\
&\!\!\!\!&\!\!\!\! |(\mathbf{\bar{h}}_{l}+\mathbf{e}_{l})^{H}\mathbf{w}|^{2} \!\geq\! E, ~\forall l. \label{eq:energy_constraint_full} 
\end{eqnarray}
\end{subequations} }
The problem \eqref{eq:robust_power_min_ori_full_channel_uncertainty}   is not convex in terms of channel uncertainties, and cannot be solved directly,
thus, we first consider the reformulation of the secrecy rate constraint \eqref{eq:sec_rate_constraint_full}, which can be relaxed as  
{\small \begin{equation}
\begin{cases} \label{eq:sec_rate_constraint_SCA_approximation}
\frac{1}{\sigma_{s}}\bigg(\mathbf{w}^{H}\mathbf{\bar{h}}_{s} \!-\! \varepsilon_{s}\|\mathbf{w}\|\bigg) \geq \sqrt{t_{2}}, \\
 \left[\!\!\begin{array}{cc}
\frac{2^{\frac{R}{2}}}{\sigma_{e}}(\mathbf{\bar{h}}_{e,k}\!+\!\mathbf{e}_{e,k})^{H}\mathbf{w} \!&\! (2^{R}\!-\!1)^{\frac{1}{2}}
\end{array}
\!\!\right]\!\!
\left[\!\!\begin{array}{cc}
\frac{2^{\frac{R}{2}}}{\sigma_{e}}\mathbf{w}^{H}(\mathbf{\bar{h}}_{e,k}\!+\!\mathbf{e}_{e,k}) \\ (2^{R}\!-\!1)^{\frac{1}{2}}
\end{array}
\!\!\right] \!\leq\! t_{2},
\end{cases}
\end{equation} }
The first constraint in \eqref{eq:sec_rate_constraint_SCA_approximation} is rewritten based on first-order Taylor approximation as follows:
{\small \begin{eqnarray}
\frac{1}{\sigma_{s}}\Re\{\mathbf{w}^{H}\mathbf{\bar{h}}_{s}\} \!-\! \frac{\varepsilon_{s}}{\sigma_{s}}\|\mathbf{w}\| \geq f^{(n)}(t_{2}),
\end{eqnarray} }
where $ f^{(n)}(t_{2}) = \sqrt{t_{2}^{(n)}} + \frac{1}{2 \sqrt{t_{2}^{(n)}}}(t_{2} - t_{2}^{(n)}) $.
\begin{lemma}\label{lemma:Nemirovski_lemma_application}
The second constraint in \eqref{eq:sec_rate_constraint_SCA_approximation} can be reformulated as
 {\small \begin{eqnarray}\label{eq:sec_rate_constraint_SDP_full_channel_uncertainty}
\mathbf{\bar{S}}_{k} \!=\! \left[\!\!\begin{array}{cc}
\mathbf{S}_{k,1} \!-\! \lambda_{k}\left[\!\!\begin{array}{cc}
\mathbf{0} \!&\! -1
\end{array}\!\!\right] \left[\!\!\begin{array}{cc}
\mathbf{0} \\ -1
\end{array}\!\!\right] \!\!&\!\! -\varepsilon_{e,k}\left[\!\!\begin{array}{cc}
\frac{2^{\frac{R}{2}}}{\sigma_{e}}\mathbf{w}^{H}  \!\\\! \mathbf{0} \!\\\! \mathbf{0}
\end{array}
\!\!\right] \\ 
 -\varepsilon_{e,k} \left[\!\!\begin{array}{ccc}
\frac{2^{\frac{R}{2}}}{\sigma_{e}}\mathbf{w}^{H}  \!&\! \mathbf{0} \!&\! \mathbf{0}
\end{array}
\!\!\right]   \!\!&\!\!  \lambda_{k}\mathbf{I}
\end{array}\!\!\right] \!\succeq\! \mathbf{0}, \forall k. \!\!\!\!\!\!\!\!\!\!\!\!\!\!\!\!\!\!\nonumber\\
\end{eqnarray}  }
where 
{\small \begin{eqnarray}
\mathbf{S}_{k,1} = \left[\!\!\begin{array}{cc}
 f^{(n)}(t_{2})\mathbf{I} \!\!&\!\! \left[\begin{array}{cc}
\frac{2^{\frac{R}{2}}}{\sigma_{e}}\mathbf{w}^{H}\mathbf{\bar{h}}_{e,k} \\ (2^{R} - 1)^{\frac{1}{2}}
 \end{array}
\right]   \\ 
 \left[\begin{array}{cc}
\frac{2^{\frac{R}{2}}}{\sigma_{e}}\mathbf{w}^{H}\mathbf{\bar{h}}_{e,k} \\ (2^{R} - 1)^{\frac{1}{2}}
 \end{array}
\right]^{H} \!\!&\!\! f^{(n)}(t_{2})
\end{array} 
 \!\!\right].
\end{eqnarray} }
\end{lemma}
\begin{IEEEproof}
Please refer to Appendix \ref{appendix:proof_of_Nemirovski_lemma}.
\end{IEEEproof}
Thus, the robust power minimization problem can be written as 
{\small \begin{eqnarray}\label{eq:robust_power_min_taylor_approximation_full}
\min_{s_{2},\mathbf{w},\lambda_{k}} &\!\!\!\!&\!\!\!\! s_{2}, \nonumber\\
s.t. &\!\!\!\!&\!\!\!\! \left[\begin{array}{cc}
s_{2} \\ \mathbf{w}
\end{array}\right] \succeq_{K} \mathbf{0}, \nonumber\\  &\!\!\!\!&\!\!\!\! \mathbf{\bar{S}}_{k}(\lambda_{k},f^{(n)}(t_{2})) \succeq \mathbf{0}, \nonumber\\ 
&\!\!\!\!&\!\!\!\! \frac{1}{\sigma_{s}}\mathbf{w}^{H}\mathbf{\bar{h}}_{s} \!-\! \frac{\varepsilon_{s}}{\sigma_{s}}\|\mathbf{w}\| \geq f^{(n)}(t_{2}), 
~\forall k, \nonumber\\ 
&\!\!\!\!&\!\!\!\!  \Re\{\mathbf{\bar{h}}_{l}^{H}\mathbf{w}\} \geq E^{\frac{1}{2}} + \varepsilon_{l} \|\mathbf{w}\|_{2},~ \Im\{\mathbf{\bar{h}}_{l}^{H}\mathbf{w}\} = 0, ~\forall l.
\end{eqnarray} }
The above problem is convex for a given $ t_{2}^{(n)} $ at each iteration. Thus, an initialization of $ t_{2} $ is given to solve the problem \eqref{eq:robust_power_min_taylor_approximation_full} by using interior-point method, which is updated iteratively. It is easily observed that $ t_{2} $ is updated to $ t_{2}^{(n+1)} = t_{2}^{n} $ when the algorithm converges. 
\subsection{Robust Harvested Energy Maximization with Channel Uncertainty}
Now, we turn our attention to robust harvested energy maximization framework, where we consider maximizing the harvested energy subject to the achieved secrecy rate and the transmit power constraints incorporating channel uncertainties. This optimization framework is formulated as  
{\small \begin{subequations}\label{eq:Robust_energy_max_ori}
\begin{eqnarray}
\max_{\mathbf{w}} \min_{\mathbf{e}_{l}} &\!\!\!\!&\!\!\!\! |(\mathbf{\bar{h}}_{l} + \mathbf{e}_{l})^{H}\mathbf{w}|^{2}, \nonumber\\
s.t. &&\!\!\!\!\!\!\!\!\! \min_{\mathbf{e}_{s}} \log\bigg( 1 \!+\! \frac{|(\mathbf{\bar{h}}_{s} + \mathbf{e}_{s})^{H}\mathbf{w}|^{2}}{\sigma_{s}^{2}} \bigg) \nonumber\\ 
&&\!\!\!\!\!\! \!-\! \max_{\mathbf{e}_{e,k}} \log\bigg(\! 1 \!+\! \frac{|(\mathbf{\bar{h}}_{e,k} \!+\! \mathbf{e}_{e,k})^{H}\mathbf{w}|^{2}}{\sigma_{e}^{2}} \!\bigg) \!\geq\! R, ~ \forall k, \label{eq:Robust_sec_rate_constraint_without_AN}\\
&\!\!\!\!&\!\!\!\! \|\mathbf{w}\|^{2} \leq P, ~\forall l.
\end{eqnarray}
\end{subequations} }
The above problem is not convex in terms of channel uncertainties and nonconvexity of the secrecy rate constraint, and cannot be solved directly. Based on the equivalent reformulations, we propose a two-level SCA based optimization algorithm to obtain the robust energy maximization problem \eqref{eq:Robust_energy_max_ori}. We denote $ \mathbf{w}^{*} $ as the optimal solution to the problem \eqref{eq:Robust_energy_max_ori}, and define $ \frac{|(\mathbf{\bar{h}}_{s} + \mathbf{e}_{s})^{H}\mathbf{w}^{*}|^{2}}{\sigma_{s}^{2}} = \tau^{*} $, then $ \mathbf{w} $ is also optimal solution of the following problem with $ \tau = \tau^{*} $. By employing a slack variable $ \tau $, we have  
{\small \begin{subequations}\label{eq:Robust_energy_max_relax_ori}
\begin{eqnarray}
\max_{\mathbf{w}} &\!\!\!\!&\!\!\!\! \min_{\mathbf{e}_{l}} |(\mathbf{\bar{h}}_{l} + \mathbf{e}_{l})^{H}\mathbf{w}|^{2}, \nonumber\\ 
s.t. &\!\!\!\!&\!\!\!\! 2^{R}\frac{|(\mathbf{\bar{h}}_{e,k} + \mathbf{e}_{e,k})^{H}\mathbf{w}|^{2}}{\sigma_{e}^{2}} + (2^{R} - 1) \leq \tau, ~\forall k, \label{eq:Eve_constraint_relax}\\ 
&\!\!\!\!&\!\!\!\! \frac{|(\mathbf{\bar{h}}_{s} + \mathbf{e}_{s})^{H}\mathbf{w}|^{2}}{\sigma_{s}^{2}} \!\geq\! \tau,  \label{eq:User_constraint_relax}\\
&\!\!\!\!&\!\!\!\! \|\mathbf{w}\|^{2} \leq P, \label{eq:Power_constraint_relax}
\end{eqnarray}
\end{subequations} }
The proof for statement is easy and the details are omitted for brevity.
Then, we consider to denote the optimal value of the problem \eqref{eq:Robust_energy_max_LMI} as $ f(\tau) $, which is a function of $ \tau $, then the following theorem holds,\\
\begin{theorem}\label{lemma:f(tau)}
The problem \eqref{eq:Robust_energy_max_ori} is equivalent to the following problem:
{\small \begin{eqnarray}\label{eq:Robust_energy_max_equivalent_optimal}
\max_{\tau \geq 0} f(\tau)
\end{eqnarray} }
\end{theorem} 
\begin{IEEEproof}
Please refer to Appendix \ref{appendix:f(tau)}.
\end{IEEEproof}
Based on \emph{Theorem 1}, the problem \eqref{eq:Robust_energy_max_equivalent_optimal} can be solved instead of solving \eqref{eq:Robust_energy_max_ori}.  Thus, we propose a two-level optimization algorithm to obtain the optimal value of $ \tau $.  In this following, we focus on reformulation of the problem \eqref{eq:Robust_energy_max_relax_ori} based on a given $ \tau $, which can rewritten by employing a slack variable for the objective function of \eqref{eq:Robust_energy_max_relax_ori} as follows:
{\small \begin{eqnarray}
\max_{\mathbf{w},t_{3}} &\!\!\!\!&\!\!\!\! t_{3}, \nonumber\\
s.t. &\!\!\!\!&\!\!\!\! |(\mathbf{\bar{h}}_{l} + \mathbf{e}_{l})^{H}\mathbf{w}|^{2} \geq t_{3}, ~\forall l, \nonumber\\ 
&\!\!\!\!&\!\!\!\! 2^{R}\frac{|(\mathbf{\bar{h}}_{e,k} + \mathbf{e}_{e,k})^{H}\mathbf{w}|^{2}}{\sigma_{e}^{2}} + (2^{R} - 1) \leq \tau, ~\forall k, \nonumber\\ 
&\!\!\!\!&\!\!\!\! \frac{|(\mathbf{\bar{h}}_{s} + \mathbf{e}_{s})^{H}\mathbf{w}|^{2}}{\sigma_{s}^{2}} \geq \tau, 
 ~\forall k, \nonumber\\
&\!\!\!\!&\!\!\!\! \|\mathbf{w}\|^{2} \leq P, 
\end{eqnarray} }
Energy constraint can be approximated based on SCA as 
{\small \begin{eqnarray}\label{eq:Robust_energy_max_energy_constraint_SCA}
\Re\{ \mathbf{w}^{H}\mathbf{\bar{h}}_{l} \} - \mathbf{e}_{l}\|\mathbf{w}\|_{2} \geq \sqrt{t_{3}^{(n)}} + \frac{1}{2\sqrt{t_{3}^{(n)}}} (t_3 - t_{3}^{(n)}), \forall l. 
\end{eqnarray}}
The secrecy rate constraint can be reformulated as 
{\small \begin{subequations}\label{eq:Robust_energy_max_sec_rate_constraint_relax}
\begin{eqnarray}
&& \left\|\! \left[\!\begin{array}{cc}
\frac{2^{\frac{R}{2}}}{\sigma_{e}}\mathbf{w}^{H}(\mathbf{\bar{h}}_{e,k} + \mathbf{e}_{e,k}) \!\!\\\!\! (2^{R} - 1)^{\frac{1}{2}}
\end{array}
  \!\right] \!\right\|_{2} \!\leq\! \sqrt{\tau},~ \forall k, \label{eq:Robust_energy_max_sec_rate_constraint_relax_01}\\
&& \frac{1}{\sigma_{s}}(\Re\{\mathbf{w}^{H}\mathbf{\bar{h}}_{s}\} - \varepsilon_{s}\|\mathbf{w}\|) \geq \sqrt{\tau}.\label{eq:Robust_energy_max_sec_rate_constraint_relax_02}
\end{eqnarray}
\end{subequations} }

From \emph{Nemirovski lemma} shown in Appendix \ref{appendix:proof_of_Nemirovski_lemma}, the constraint \eqref{eq:Robust_energy_max_sec_rate_constraint_relax_01} can be written as 
{\small \begin{eqnarray}\label{eq:Robust_energy_max_sec_rate_constraint_relax_01_LMI}
\mathbf{\bar{A}}_{k} \!=\! \left[\!\! \begin{array}{cc}
\mathbf{A}_{k} \!-\! \mu_{k}\left[\!\!\begin{array}{cc}
\mathbf{0} \!\!&\!\! -1
\end{array} \!\!\right] \left[\!\!\begin{array}{cc}
\mathbf{0} \!\!\\\!\! -1
\end{array} \!\!\right]   \!\!&\!\! -\varepsilon_{e,k} \left[\!\! \begin{array}{cc}
\frac{2^{\frac{R}{2}}}{\sigma_{e}}\mathbf{w}^{H} \\ \mathbf{0} \!\!\\\!\! \mathbf{0}
\end{array}
  \!\!\right]  \!\!\\\!\! -\varepsilon_{e,k} \left[\!\! \begin{array}{ccc}
\frac{2^{\frac{R}{2}}}{\sigma_{e}}\mathbf{w}^{H} \!\!&\!\! \mathbf{0} & \mathbf{0}
\end{array}
  \!\!\right]
   \!\!&\!\! \mu_{k}\mathbf{I}
\end{array}
 \!\!\right] \succeq \mathbf{0}, 
\end{eqnarray} }
where 
{\small \begin{eqnarray}
\mathbf{A}_{k} = \left[\begin{array}{cc}
\sqrt{\tau}\mathbf{I}  & \left[\begin{array}{cc}
\frac{2^{\frac{R}{2}}}{\sigma_{e}}\mathbf{w}^{H}\mathbf{\bar{h}}_{e,k}  \\ (2^{R} - 1)^{\frac{1}{2}}
\end{array}
\right] \\ \left[\begin{array}{cc}
\frac{2^{\frac{R}{2}}}{\sigma_{e}}\mathbf{w}^{H}\mathbf{\bar{h}}_{e,k}  \\ (2^{R} - 1)^{\frac{1}{2}}
\end{array}
\right]^{H}
  & \sqrt{\tau}
\end{array}
\right]
\end{eqnarray} }
Thus, the robust harvested energy maximization problem is equivalently reformulated as 
{\small \begin{eqnarray}\label{eq:Robust_energy_max_LMI}
\max_{\mathbf{w}, t_{3}} &\!\!\!\!&\!\!\!\! t_{3} \nonumber\\ 
s.t. &\!\!\!\!&\!\!\!\! \|\mathbf{w}\|^{2} \leq P, \nonumber\\ 
 &\!\!\!\!&\!\!\!\! \eqref{eq:Robust_energy_max_energy_constraint_SCA}, ~ \eqref{eq:Robust_energy_max_sec_rate_constraint_relax_01_LMI}, ~\eqref{eq:Robust_energy_max_sec_rate_constraint_relax_02}.
\end{eqnarray} }
The problem \eqref{eq:Robust_energy_max_LMI} is convex for a given $ \tau $ and a initialized value of $ t_{3}^{(n)} $, the SCA algorithm also can be employed to approximate $ t_{3}^{(n)} $ for \eqref{eq:Robust_energy_max_sec_rate_constraint_relax_02},
whereas we propose a straightforward line based search algorithm to obtain optimal $ \tau $. In order to carry out this algorithm, we will determine the iterative and feasible region of $ \tau $ (i.e., $ \tau_{min} \leq \tau \leq \tau_{max} $), where the optimal value of $ \tau^{*} $ is inside this region. According to the problem \eqref{eq:Robust_energy_max_equivalent_optimal}, we can easily determine the lower bound of $ \tau_{min} = 0 $. Then, we will show the upper bound $ \tau_{max} $ by using inequality properties:
{\small \begin{eqnarray}\label{eq:upper_bound_derived}
\frac{|(\mathbf{\bar{h}}_{s} + \mathbf{e}_{s})^{H}\mathbf{w}|^{2}}{\sigma_{s}^{2}} &\!\!\!\!\leq&\!\!\!\! \frac{\| \mathbf{w} \|^{2}}{\sigma_{s}^{2}} \|\mathbf{\bar{h}}_{s} + \mathbf{e}_{s}\|^{2} \nonumber\\ 
&\!\!\!\!\leq&\!\!\!\! \frac{P}{\sigma_{s}^{2}} (\| \mathbf{\bar{h}}_{s} \|_{2} + \| \mathbf{e}_{s} \|_{2})^{2} \nonumber\\ 
&\!\!\!\!\leq&\!\!\!\!  \frac{P}{\sigma_{s}^{2}} (\|\mathbf{\bar{h}}_{s}\|_{2} + \varepsilon_{s})^{2}.
\end{eqnarray}  }
The upper bound in \eqref{eq:upper_bound_derived} can be derived based on \emph{Cauchy-schwarz} and triangle inequalities, as well as $ \|\mathbf{w}\|^{2} \leq P $ and $ \|\mathbf{e}_{s}\| \leq \varepsilon_{s} $. Hence, the iterative region can be obtain as $ 0 \leq \tau \leq \frac{P}{\sigma_{s}^{2}} (\|\mathbf{\bar{h}}_{s}\|_{2} + \varepsilon_{s})^{2} $. From the reformulation for the original problem \eqref{eq:Robust_energy_max_ori} and iterative region, we summarize one-dimensional search algorithm as Table \ref{Table:SCA_based_one_dimension_algorithm}.
{\small \begin{table}
\vspace{-0.05in}
\hrule
\caption{\label{Table:SCA_based_one_dimension_algorithm} One-Dimensional Search for \eqref{eq:Robust_energy_max_LMI}}
\vspace{0.05in}
\hrule 
\vspace{0.01in}
\begin{enumerate}
   \item Given $ \tau_{max} $, $ \Delta{\tau} $, and an initialization value of $ t_{3}^{(n)} = t_{3}^{(0)} $.
  \item \textbf{Outer Iteration loop begin}\\
  \textbf{If} $ \tau \leq \tau_{\max} $, then  
       \begin{enumerate}
        \vspace{-1mm}     
        \item \textbf{Inner Iteration loop begin}\\
        \textbf{If} the problem \eqref{eq:Robust_energy_max_LMI} is feasible, then 
        \begin{enumerate}
        \item Solving \eqref{eq:Robust_energy_max_LMI} based on SCA algorithm for a initialization $ t_{3}^{n} $ based on a given $ \tau $. 
        \item Set $ t_{3} = t_{3}^{(n)} $.
        \end{enumerate}
        \textbf{else} \\
        $~~~~~$\textbf{Break}\\
       \textbf{end}
       \item \textbf{Inner Iteration loop end until the SCA algorithm converges.}
        \item Updating $ \tau = \tau + \Delta{\tau} $.
      \end{enumerate}
 $~$\textbf{end}
        \item \textbf{Iteration loop end until the required accuracy.}
        \item Obtain $ \tau^{*} $ by solving a sequence of problems \eqref{eq:Robust_energy_max_equivalent_optimal} for a given $ \tau $, and the optimal transmit beamforming $ \mathbf{w}^{*} $ can be obtained. 
\end{enumerate}
\hrule
\vspace{0.1in}
\end{table} }

\section{Secure SWIPT of Joint Beamforming and Artificial Noise Design}\label{section Secure_SWIPT_of_Joint_Beamforming_and_Artificial_Noise_Design}
In the previous section, we solved the secrecy rate optimization problems to optimize the transmit beamforming without AN, where the proposed algorithms are implemented without rank-relaxation  at the expense of high computational complexity. In this section, we extend our attention to joint signal beamforming and artificial noise (AN) design in secrecy rate optimization problem, where the legitimate transmitter sends the signal with AN in order to introduce more interferences to the eavesdroppers. In addition, AN can be harvested with signal beamforming by the EH receivers, which improves the performance of SWIPT. The secrecy rate optimization problem can be formulated into SDP. Unlike \cite{Tian_Maoxin_SPL_J15}, where it has shown that the rank of the optimal solution obtained from line search algorithm was less than or equal to 2. In this paper, we provide a novel SDP relaxation for the secrecy rate optimization problem, and exactly show the optimal solution always returns rank-one. 
\subsection{Problem Formulation}
We first write the secrecy rate maximization problem with the transmit power and the minimum EH constraints, where the secured transmit beamforming and AN are jointly designed. This optimization problem can be formulated as  
{\small \begin{eqnarray}\label{eq:sec_rate_max_with_AN_ori}
\max_{\mathbf{Q},\mathbf{W}} && \min_{k} R_{s} \!-\! R_{e,k}, ~\forall k, \nonumber\\
s.t. && \textrm{Tr}(\mathbf{Q}_{s} + \mathbf{W}) \leq P, 
~~ [\mathbf{Q}_{s} + \mathbf{W}]_{(i,i)} \leq p_{i},~\forall i,
 \nonumber\\
&& \min_{l} \mathbf{h}_{l}^{H}(\mathbf{Q}_{s} + \mathbf{W})\mathbf{h}_{l} \geq E_{l}, ~\forall l.  
\end{eqnarray} }
where $ \mathbf{Q}_{s} = \mathbb{E}\{ \mathbf{w}\mathbf{w}^{H} \} $ is transmit covariance matrix, $ \mathbf{W} = \mathbb{E}\{ \mathbf{v}\mathbf{v}^{H} \} $ is AN beamforming, $ [\mathbf{Q}_{s} + \mathbf{W}]_{(i,i)} $ ($ i = 1,..., N_{T} $) represents each antenna transmit power constraint, and the mutual information at the legitimate user and $ k $-th eavesdropper can be written as 
{\small \begin{eqnarray}
R_{s} &\!\!\!\!\!=&\!\!\!\!\! \log\bigg(1+\frac{\mathbf{h}_{s}^{H}\mathbf{Q}_{s}\mathbf{h}_{s}^{H}}{\mathbf{h}_{s}^{H}\mathbf{W}\mathbf{h}_{s}^{H} + \sigma_{s}^{2}}\bigg), \nonumber\\
R_{e,k} &\!\!\!\!\!=&\!\!\!\!\! \log\bigg(1+\frac{\mathbf{h}_{e,k}^{H}\mathbf{Q}_{s}\mathbf{h}_{e,k}^{H}}{\mathbf{h}_{e,k}^{H}\mathbf{W}\mathbf{h}_{e,k}^{H} + \sigma_{e}^{2}}\bigg). \nonumber
\end{eqnarray} }
\emph{Remark 1}: The relaxation method to be proposed is also suitable for the scenario that the eavesdroppers and the EH receivers are equipped with multiple antennas. We will provide the analysis in Section \ref{subsection_multiple_antennas_case_for_Eves_and_EH_Rx}.
\subsection{Secrecy Rate Optimization}
For the secrecy rate maximization problem \eqref{eq:sec_rate_max_with_AN_ori}, we will provide two different methods to joint optimize transmit covariance matrix and AN beamforming: 1) SDP-based one-dimensional line search, 2) SDP based on SCA.  
\subsubsection{SDP-Based One-Dimensional Line Search Method}\label{subsubsection SDP_Based_One_Dimensional_Line_Search_Method} 
We first introduce a slack variable $ t $, and the problem \eqref{eq:sec_rate_max_with_AN_ori} is rewritten as 
{\small \begin{subequations}\label{eq:Sec_rate_max_pro_slack_variable}
\begin{eqnarray}
\max_{\mathbf{Q}_{s},\mathbf{W},t} &\!\!\!\!&\!\!\!\! R_{s} + \log(t), \nonumber\\
s.t. &&\!\!\!\!\!\!\!\! \log\bigg(1+\frac{\mathbf{h}_{e,k}^{H}\mathbf{Q}_{s}\mathbf{h}_{e,k}^{H}}{\mathbf{h}_{e,k}^{H}\mathbf{W}\mathbf{h}_{e,k}^{H} + \sigma_{e}^{2}}\bigg) \leq \log(\frac{1}{t}), ~ \forall k, \label{eq:Eve_rate_slack_variable}\\
&&\!\!\!\!\!\!\!\! \textrm{Tr}(\mathbf{Q}_{s} + \mathbf{W}) \leq P, 
~~ \textrm{Tr}[\mathbf{A}_{i}(\mathbf{Q}_{s} + \mathbf{W})] \leq p_{i},  ~\forall i, \label{eq:Power_constraints}\\
&&\!\!\!\!\!\!\!\! \mathbf{h}_{l}^{H}(\mathbf{Q}_{s} + \mathbf{W})\mathbf{h}_{l} \geq E_{l}, ~\forall l, \label{eq:EH_constraint_other}\\
&&\!\!\!\!\!\!\!\! \mathbf{Q}_{s} \succeq \mathbf{0}, \mathbf{W} \succeq \mathbf{0}, t \geq 0,  \nonumber
\end{eqnarray}
\end{subequations} }
where $ \mathbf{A}_{i} = \mathbf{a}_{i}\mathbf{a}_{i}^{H} $ is given antenna design parameters to adjust each antenna power budget, and $ \mathbf{a}_{i} $ is unit $ i $-th vector (i.e., $ [\mathbf{a}_{i}]_{j} = 1 $ for $ i = j $ and $ [\mathbf{a}_{i}]_{j} = 0 $ for $ i \neq j $). The specific applications of per-antenna power constraint have been already described in \cite{Ma_TSP_J13,Khandaker_TIFS_J15}.
The problem \eqref{eq:Sec_rate_max_pro_slack_variable} is still not convex in terms of the constraint \eqref{eq:Eve_rate_slack_variable}, and cannot be solved directly, thus, it can be formulated as a two-stage optimization problem, the outer problem is a function of $ t $, which can be written as  
{\small \begin{eqnarray}
\max_{t} && \log(1+ f(t)) + \log(t), \nonumber\\
s.t. && t_{\min} \leq t \leq 1, 
\end{eqnarray} }
The lower bound $ t_{\min} $ can be derived as follows:
{\small \begin{eqnarray} 
t &\!\!\!\! \geq &\!\!\!\! \bigg(1 + \frac{\mathbf{h}_{s}^{H}\mathbf{Q}_{s}\mathbf{h}_{s}^{H}}{\mathbf{h}_{s}^{H}\mathbf{W}\mathbf{h}_{s}^{H} + \sigma_{s}^{2}}\bigg)^{-1} \! \geq \! \bigg(1 + \frac{\mathbf{h}_{s}^{H}\mathbf{Q}_{s}\mathbf{h}_{s}^{H}}{\sigma_{s}^{2}}\bigg)^{-1} \nonumber\\
&\!\!\!\! \geq &\!\!\!\! \bigg(1 + \frac{\lambda_{\max}(\mathbf{Q}_{s})\|\mathbf{h}_{s}\|^{2}}{\sigma_{s}^{2}}\bigg)^{-1} \! \geq \! \bigg(1 + \frac{\textrm{Tr}(\mathbf{Q}_{s})\|\mathbf{h}_{s}\|^{2}}{\sigma_{s}^{2}}\bigg)^{-1} \nonumber\\
&\!\!\!\! \geq &\!\!\!\! \bigg( 1 + \frac{P \|\mathbf{h}_{s}\|^{2}}{\sigma_{s}^{2}} \bigg)^{-1} \!=\! t_{\min} .
\end{eqnarray} }
Then, the inner problem can recast for a given $ t $ as follows:
{\small \begin{eqnarray}
f(t) = \max_{\mathbf{Q}_{s},\mathbf{W},t} && \frac{\mathbf{h}_{s}^{H}\mathbf{Q}_{s}\mathbf{h}_{s}^{H}}{\mathbf{h}_{s}^{H}\mathbf{W}\mathbf{h}_{s}^{H} + \sigma_{s}^{2}}, \nonumber\\
s.t. && \eqref{eq:Eve_rate_slack_variable}-\eqref{eq:EH_constraint_other}.
\end{eqnarray} }
It easily verified that the constraint \eqref{eq:Sec_rate_max_pro_slack_variable} can be reformulated as 
{\small \begin{eqnarray}\label{eq:Eve_rate_constraint_reformulation}
\mathbf{h}_{e,k}^{H}\bigg[\mathbf{Q}_{s} \!-\! (\frac{1}{t} \!-\! 1)\mathbf{W}\bigg]\mathbf{h}_{e,k} \!\leq\! (\frac{1}{t} \!-\! 1)\sigma_{e}^{2}, 
\end{eqnarray} }
Thus the problem can recast for a given $ t $ as follows:
{\small \begin{eqnarray}\label{eq:Inner_sec_rate_max_quasi_convex_problem}
f(t) = \max_{\mathbf{Q}_{s},\mathbf{W},t} && \frac{\mathbf{h}_{s}^{H}\mathbf{Q}_{s}\mathbf{h}_{s}^{H}}{\mathbf{h}_{s}^{H}\mathbf{W}\mathbf{h}_{s}^{H} + \sigma_{s}^{2}}, \nonumber\\
s.t. && \eqref{eq:Eve_rate_constraint_reformulation},~ 
 \eqref{eq:Power_constraints},~\eqref{eq:EH_constraint_other}.
\end{eqnarray} }
The problem \eqref{eq:Inner_sec_rate_max_quasi_convex_problem} is a quasi-convex problem, thus we consider Charnes-Cooper transformation to convert it into a convex problem. Let us introduce $ \delta $ so that the following relations hold:
{\small \begin{eqnarray}\label{eq:CC_transformation}
\mathbf{Q}_{s} = \frac{\mathbf{\bar{Q}}_{s}}{\delta},~ \mathbf{W} = \frac{\mathbf{\bar{W}}}{\delta}
\end{eqnarray} }
Thus, we have 
{\small \begin{eqnarray}\label{eq:Sec_rate_max_with_AN_perfect_result}
f(t)\! =\! \max_{\mathbf{\bar{Q}}_{s},\mathbf{\bar{W}}} &\!\!\!\!\!\!&\!\!\!\!\!\! \mathbf{h}_{s}^{H}\mathbf{\bar{Q}}_{s}\mathbf{h}_{s}, \nonumber\\
 s.t. &&\!\!\!\!\!\!\!\!\!\! \mathbf{h}_{s}^{H}\mathbf{\bar{W}}\mathbf{h}_{s} + \delta\sigma_{b}^{2} = 1, \nonumber\\
&&\!\!\!\!\!\!\!\!\!\!\!\!\!\!\!\!\!\! \mathbf{h}_{e,k}^{H}\bigg[\mathbf{\bar{Q}}_{s} \!-\! (\frac{1}{t} \!-\! 1)\mathbf{\bar{W}}\bigg]\mathbf{h}_{e,k} \!\leq\! (\frac{1}{t} \!-\! 1)\delta\sigma_{e}^{2}, \nonumber\\
&&\!\!\!\!\!\!\!\!\!\!\!\!\!\!\!\!\!\!\!\!\!\! \textrm{Tr}(\mathbf{\bar{Q}}_{s} + \mathbf{\bar{W}}) \leq \delta P, ~\textrm{Tr}[\mathbf{A}_{i}(\mathbf{\bar{Q}}_{s} + \mathbf{\bar{W}})] \leq \delta p_{i}, ~\forall i, \nonumber\\
&&\!\!\!\!\!\!\!\!\!\!\!\!\!\!\!\!\!\!\!\!\!\! \mathbf{h}_{l}^{H}(\mathbf{\bar{Q}}_{s} + \mathbf{\bar{W}})\mathbf{h}_{l} \geq \delta E_{l},  ~\forall l, ~ \mathbf{\bar{Q}}_{s} \succeq \mathbf{0}, ~\mathbf{\bar{W}} \succeq \mathbf{0}.
\end{eqnarray} }
The problem \eqref{eq:Sec_rate_max_with_AN_perfect_result} is a convex problem, 
and can be solved efficiently by using interior-point method \cite{boyd_B04}. 
Thus, the optimal solution of the problem \eqref{eq:Inner_sec_rate_max_quasi_convex_problem} can be obtained through \eqref{eq:CC_transformation}, once the problem \eqref{eq:Sec_rate_max_with_AN_perfect_result} has been solved. 
\subsubsection{Tightness Analyses of Rank Relaxation}
Now we investigate the tightness of the rank relaxation to the problem \eqref{eq:Inner_sec_rate_max_quasi_convex_problem}, we assume that the optimal value $ f(t) $ can be obtained by the optimal solution of the problem \eqref{eq:Inner_sec_rate_max_quasi_convex_problem}, we have the following inequality, 
{\small \begin{eqnarray}\label{eq:f(t)_constraint}
\frac{\mathbf{h}_{s}^{H}\mathbf{Q}_{s}\mathbf{h}_{s}}{\mathbf{h}_{s}^{H}\mathbf{W}\mathbf{h}_{s} \!+\! \sigma_{s}^{2}} \!\geq\! f(t) \!\Rightarrow\! \mathbf{h}_{s}^{H}[\mathbf{Q}_{s} \!-\! f(t)\mathbf{W}]\mathbf{h}_{s} \!\geq\! f(t)\sigma_{s}^{2},
\end{eqnarray} }
Thus, we consider the following power minimization,
{\small \begin{eqnarray}\label{eq:power_min_to_rank_one}
\min_{\mathbf{Q}_{s}} &\!\!\!\!&\!\!\!\! \textrm{Tr}(\mathbf{Q}_{s})  \nonumber\\
s.t. &\!\!\!\!&\!\!\!\! \eqref{eq:f(t)_constraint},~ 
 \eqref{eq:Power_constraints},~\eqref{eq:EH_constraint_other}.
\end{eqnarray} }
It is easily verified that the feasible solution of problem \eqref{eq:power_min_to_rank_one} is the optimal of \eqref{eq:Inner_sec_rate_max_quasi_convex_problem} due to the constraints \eqref{eq:f(t)_constraint}, \eqref{eq:Power_constraints}, \eqref{eq:EH_constraint_other}.
 Thus, the following theorem is provided to show every optimal solution of the problem \eqref{eq:power_min_to_rank_one} is rank-one:
\begin{theorem}\label{lemma:rank-one_proof_of_sec_rate_max_with_AN}
If the problem \eqref{eq:Inner_sec_rate_max_quasi_convex_problem} is feasible, then there always exists optimal solution (i.e., $ \mathbf{Q}_{s} $) satisfies $ \textrm{rank}(\mathbf{Q}_{s}) \leq 1 $.
\end{theorem}
\begin{IEEEproof}
Please refer to Appendix \ref{appendix:rank-one_proof_of_sec_rate_max_with_AN}.
\end{IEEEproof}
From \emph{Theorem 2}, tightness analysis has been provided so that the problem \eqref{eq:Inner_sec_rate_max_quasi_convex_problem} has a rank-one solution for all feasible $ t $. 

\subsubsection{SDP Based Successive Convex Approximation}\label{subsubsection SDP_Based_SCA}
In this section, we propose a SDP based \emph{successive linear approximation} (SCA) algorithm to joint optimization for transmit beamforming and AN, thus, rewrite the problem \eqref{eq:sec_rate_max_with_AN_ori} as  
{\small \begin{subequations}\label{eq:SCA_sec_rate_max_ori}
\begin{eqnarray}
\!\!\!\!\!\!\!\!\!\!\!\! \min_{\mathbf{Q}_{s},\mathbf{W}} &&\!\!\!\!\!\!\!\!\!\! \max_{k} \frac{\bigg(\sigma_{e}^{2} \!+\! \textrm{Tr}[\mathbf{h}_{e,k}\mathbf{h}_{e,k}^{H}(\mathbf{Q}_{s} \!+\! \mathbf{W})]\bigg)\bigg( \sigma_{s}^{2} \!+\! \textrm{Tr}(\mathbf{h}_{s}\mathbf{h}_{s}^{H}\mathbf{W}) \bigg)}{\bigg(\sigma_{s}^{2} \!+\! \textrm{Tr}[\mathbf{h}_{s}\mathbf{h}_{s}^{H}(\mathbf{Q}_{s} \!+\! \mathbf{W})]\bigg)\bigg( \sigma_{e}^{2} \!+\! \textrm{Tr}(\mathbf{h}_{e,k}\mathbf{h}_{e,k}^{H}\mathbf{W}) \bigg)} \nonumber\\
s.t. &&\!\!\!\!\!\!  \textrm{Tr}(\mathbf{Q}_{s} + \mathbf{W}) \leq P, 
~~ \textrm{Tr}[\mathbf{A}_{i}(\mathbf{Q}_{s} + \mathbf{W})] \leq p_{i},  ~\forall i,
 \label{eq:power_constraints}\\
&&\!\!\!\!\!\!  \mathbf{h}_{l}^{H}(\mathbf{Q}_{s} + \mathbf{W})\mathbf{h}_{l} \geq E_{l}, ~\forall l.  \label{eq:EH_constraint}
\end{eqnarray} 
\end{subequations} }
The above problem is not convex in terms of the objective function. Let us introduce the following exponential variables to equivalently modified the objective function. 
{\small \begin{eqnarray}\label{eq:SCA_expoential_variables}
e^{x_{0}} &\!\!\!\!\!=&\!\!\!\!\!\! \sigma_{s}^{2} \!+\! \textrm{Tr}[\mathbf{h}_{s}\mathbf{h}_{s}^{H}(\mathbf{Q}_{s} \!+\! \mathbf{W})],~e^{x_{k}} \!=\! \sigma_{e}^{2} \!+\! \textrm{Tr}(\mathbf{h}_{e,k}\mathbf{h}_{e,k}^{H}\mathbf{W}), \nonumber\\
e^{y_{k}} &\!\!\!\!\!=&\!\!\!\!\!\! \sigma_{e}^{2} \!+\! \textrm{Tr}[\mathbf{h}_{e,k}\mathbf{h}_{e,k}^{H}(\mathbf{Q}_{s} \!+\! \mathbf{W})],~ e^{y_{0}} \!=\! \sigma_{s}^{2} \!+\! \textrm{Tr}(\mathbf{h}_{s}\mathbf{h}_{s}^{H}\mathbf{W}). \!\!\!\!\!\!\!\!\!\!\!\!\!\!\nonumber\\
\end{eqnarray} }
Thus, the problem \eqref{eq:SCA_sec_rate_max_ori} is rewritten by introducing a slack variable $ \tau $ as 
{\small \begin{subequations}\label{eq:SCA_sec_rate_max_slack_variables_relaxed}
\begin{eqnarray}
&\!\!\!\!\!\!\!\!&\!\!\!\!\!\!\!\! \min_{\mathbf{Q}_{s},\mathbf{W},x_{0},y_{0},x_{k},y_{k}}  \tau \label{eq:SCA_objective_function}\\
s.t. &&\!\!\!\!\!  e^{y_{0} \!-\! x_{0} \!+\! y_{k} \!-\! x_{k}} \!\leq\! \tau, \label{eq:SCA_slack_variable_constraint}\\
&&\!\!\!\!\!\!\!\!\!\!\!\!\!\!\!\!\!\!\!\! \sigma_{s}^{2} \!+\! \textrm{Tr}[\mathbf{h}_{s}\mathbf{h}_{s}^{H}(\mathbf{Q}_{s}\! +\! \mathbf{W})] \!\geq\! e^{x_{0}},~\sigma_{e}^{2} \!+\! \textrm{Tr}(\mathbf{h}_{e,k}\mathbf{h}_{e,k}^{H}\mathbf{W}) \geq e^{x_{k}}, \label{eq:SCA_constraint_01}\nonumber\\\\
&&\!\!\!\!\!\!\!\!\!\!\!\!\!\!\!\!\!\!\!\! \sigma_{e}^{2} \!+\! \textrm{Tr}[\mathbf{h}_{e,k}\mathbf{h}_{e,k}^{H}(\mathbf{Q}_{s} \!+\! \mathbf{W})] \!\leq\! e^{y_{k}}, ~
 \sigma_{s}^{2} \!+\! \textrm{Tr}(\mathbf{h}_{s}\mathbf{h}_{s}^{H}\mathbf{W}) \!\leq\! e^{y_{0}}, \label{eq:SCA_constraint_02}\nonumber\\\\
&& \eqref{eq:power_constraints},~\eqref{eq:EH_constraint}.
\end{eqnarray}
\end{subequations} }
The above problem is not still convex in terms the constraint \eqref{eq:SCA_constraint_02}. 
Thus, Taylor series expansion (i.e., $a^{\hat{x}} + a^{\hat{x}}\ln a(x - \hat{x}) \leq a^{x}$) is employed to linearise \eqref{eq:SCA_constraint_02} as follows:
{\small \begin{subequations}\label{eq:SCA_linear_approximations}
\begin{eqnarray}
\sigma_{e}^{2} \!+\! \textrm{Tr}[\mathbf{h}_{e,k}\mathbf{h}_{e,k}^{H}(\mathbf{Q}_{s} \!+\! \mathbf{W})] \leq e^{\hat{y}_{k}}(y_{k} \!-\! \hat{y}_{k} \!+\! 1), \\
 \sigma_{s}^{2} \!+\! \textrm{Tr}(\mathbf{h}_{s}\mathbf{h}_{s}^{H}\mathbf{W}) \!\leq\! e^{\hat{y}_{0}}(y_{0} \!-\! \hat{y}_{0} \!+\! 1),
\end{eqnarray}
\end{subequations} }
Thus, the secrecy rate maximization problem can be relaxed as 
{\small \begin{eqnarray}\label{eq:SCA_sec_rate_max_results}
&\!\!\!\!&\!\!\!\!\min_{\mathbf{Q}_{s},\mathbf{W},x_{0},y_{0},x_{k},y_{k},\tau}  \tau \nonumber\\ 
&&\!\!\!\!\!\!\!\!s.t. ~~ \eqref{eq:power_constraints},~\eqref{eq:EH_constraint},~\eqref{eq:SCA_slack_variable_constraint},~\eqref{eq:SCA_constraint_01},~\eqref{eq:SCA_linear_approximations}, ~\forall k.
\end{eqnarray} }

From SCA, the approximation with current optimal solution can be updated iteratively until the constraints \eqref{eq:SCA_constraint_01} and \eqref{eq:SCA_constraint_02} hold with equality, which implies the problem \eqref{eq:SCA_sec_rate_max_ori} is optimally solved. This SCA algorithm is outlined as Table 
\ref{Table:SCA_algorithm_robust_sec_rate_max_with_AN_first_new} . 
The optimal solution obtained by the SDP based SCA algorithm at $ n $-the iteration is assumed to be ($ \mathbf{Q}_{s}^{*}(n),\mathbf{W}^{*}(n),x_{0}^{*}(n),y_{0}^{*}(n),x_{k}^{*}(n),y_{k}^{*}(n),\tau^{*}(n) $), which can achieve a stable point when the SDP based SCA algorithm converges, the proof has been shown in \cite{Wei_Chiang_Li_TSP_J13}. 
Now, we consider the tightness analysis for the problem \eqref{eq:SCA_sec_rate_max_slack_variables_relaxed} due to rank relaxation. It is assumed that ($ \mathbf{Q}_{s}^{*},\mathbf{W}^{*} $) are the optimal solution of the problem \eqref{eq:SCA_sec_rate_max_ori} that are obtained by solving the problem \eqref{eq:SCA_sec_rate_max_results} with the SDP based SCA algorithm, and the corresponding slack variables (i.e., $ x_{0}^{*},y_{0}^{*},x_{k}^{*},y_{k}^{*},\tau^{*} $) can be obtained by \eqref{eq:SCA_expoential_variables} and \eqref{eq:SCA_slack_variable_constraint}, respectively. Thus, we consider the following problem:
{\small \begin{eqnarray}\label{eq:power_min_SCA_results}
\min_{\mathbf{Q}_{s},\mathbf{W}} &\!\!\!\!&\!\!\!\! \textrm{Tr}(\mathbf{Q}_{s}) \nonumber\\
s.t. 
&\!\!\!\!&\!\!\!\! \eqref{eq:power_constraints},~\eqref{eq:EH_constraint}, \nonumber\\
&&\!\!\!\!\!\!\!\!\!\!\!\!\!\!\!\!\!\!\!\!\!\!\!\! \sigma_{s}^{2} \!+\! \textrm{Tr}[\mathbf{h}_{s}\mathbf{h}_{s}^{H}(\mathbf{Q}_{s} \!+\! \mathbf{W})] \!\geq\! e^{x_{0}^{*}},~\sigma_{e}^{2} \!+\! \textrm{Tr}(\mathbf{h}_{e,k}\mathbf{h}_{e,k}^{H}\mathbf{W}) \!\geq\! e^{x_{k}^{*}}, \nonumber\\
&&\!\!\!\!\!\!\!\!\!\!\!\!\!\!\!\!\!\!\!\!\!\!\!\! \sigma_{e}^{2} \!+\! \textrm{Tr}[\mathbf{h}_{e,k}\mathbf{h}_{e,k}^{H}(\mathbf{Q}_{s} \!+\! \mathbf{W})] \!\leq\! e^{y_{k}^{*}}, ~
 \sigma_{s}^{2} \!+\! \textrm{Tr}(\mathbf{h}_{s}\mathbf{h}_{s}^{H}\mathbf{W}) \!\leq\! e^{y_{0}^{*}}, ~\forall k. \nonumber\\\!\!\!\!\!\!\!\!\!
\end{eqnarray} }
We also assume that the optimal solutions of the above problem can be denoted as ($ \mathbf{\hat{Q}}_{s},\mathbf{\hat{W}} $), which are the feasible solution of the problem \eqref{eq:SCA_sec_rate_max_ori}. Hence, the objective value $ \hat{\tau} $ is obtained by substituting ($ \mathbf{\hat{Q}}_{s},\mathbf{\hat{W}} $) into \eqref{eq:SCA_sec_rate_max_ori}, and we have $ \hat{\tau} \leq \tau^{*} $, which implies ($ \mathbf{\hat{Q}}_{s},\mathbf{\hat{W}} $) is at least the same optimal solution to ($ \mathbf{Q}_{s}^{*},\mathbf{W}^{*} $) for \eqref{eq:SCA_sec_rate_max_ori}. 
Thus, provided the problem \eqref{eq:SCA_sec_rate_max_ori} is feasible for positive secrecy rate, there always exists the optimal solution is rank-one, the proof is similar to that of \emph{Theorem} \ref{lemma:rank-one_proof_of_sec_rate_max_with_AN}.

{\small \begin{table}
\vspace{-0.05in}
\hrule
\caption{\label{Table:SCA_algorithm_robust_sec_rate_max_with_AN_first_new} SCA algorithm for the robust secrecy rate maximization problem \eqref{eq:SCA_sec_rate_max_slack_variables_relaxed}.}
\vspace{0.05in}
\hrule 
\vspace{0.01in}
\begin{enumerate}
   \item Initialize ($ \mathbf{Q}_{s}[0],\mathbf{W}[0] $) so that \eqref{eq:SCA_sec_rate_max_slack_variables_relaxed} is feasible, and given $ \kappa $ as the tolerance factor for stopping criterion.
  \item \textbf{Iteration loop begin}: 
       \begin{enumerate}
        \vspace{-1mm}  
        \item Updating ($ x_{0}[n],x_{k}[n],y_{0}[n],y_{k}[n] $) by \eqref{eq:SCA_expoential_variables}. 
        \item Solving \eqref{eq:SCA_sec_rate_max_results} with ($ x_{0}[n],x_{k}[k],y_{0}[n],y_{k}[n] $) to obtain ($ \mathbf{Q}_{s}[n],\mathbf{W}[n] $).
        \end{enumerate}
         \item \textbf{Iteration loop end until stopping criterion $ |\tau(n+1) - \tau(n)| \leq \kappa $.}      
\end{enumerate}        
\hrule
\vspace{0.1in}
\end{table} }
\vspace{-1em}
\subsection{Robust Secrecy Rate Optimization}
In the previous section, we have solved secrecy rate maximization problem based on the assumption that the legitimate transmitter has perfect CSI of the legitimate user, the eavesdroppers and the EH receivers. However, it is not always possible to have perfect CSI at the transmitter due to quantization errors and channel estimation. Therefore, robust secrecy rate optimization is employed to jointly optimize the transmit beamforming and AN by incorporating channel uncertainties, which have been shown in Section \ref{subsection channel uncertainty}. In addition, robust power constraint per antenna is considered, where the Hermitian positive semidefinite (PSD) matrix $ \mathbf{A}_{i} $ is not available at the legitimate transmitter, thus the true PSD matrix can be written as   
{\small \begin{eqnarray}
\mathbf{A}_{i} = \mathbf{\bar{A}}_{i} + \mathbf{\Delta}_{i},~ \|\mathbf{\Delta}_{i}\|_{F} \leq \varepsilon_{i}, ~\forall i,
\end{eqnarray}}
where $ \mathbf{\bar{A}}_{i} \in \mathbb{H}_{+}^{N_{T}} $ is the estimated Hermitian PSD matrix, $ \mathbf{\Delta}_{i} $ is estimated error of the matrix $ \mathbf{\bar{A}}_{i} $, which can be modelled as a spherical uncertainty with a norm-bound $ \varepsilon_{i} $. In the following, we consider one dimensional search and successive convex optimization methods to solve the robust secrecy rate maximization problem by incorporating with the channel uncertainties.
\subsubsection{Robust SDP Based One Dimension Search}\label{subsection robust_SDP_based_one_diemension_search}
In this subsection, robust SDP based one-dimensional search algorithm is proposed to jointly optimize transmit beamforming and AN by incorporating with channel uncertainties. Thus, the two-level optimization framework have been discussed in the previous subsection is employed by incorporating with channel uncertainties. Since the outer problem does not involve the channel uncertainties that is similar to Section \ref{subsubsection SDP_Based_One_Dimensional_Line_Search_Method}, thus we focus on the inner problem for a given $ t $, which can be reformulated as  
{\small \begin{eqnarray}\label{eq:eq:Robust_sec_rate_relaxed_00}
&\!\!\!\!&\!\!\!\! f(t) \!=\! \max_{\mathbf{Q}_{s},\mathbf{W},t}  ~\frac{(\mathbf{\bar{h}}_{s} \!+\! \mathbf{e}_{s})^{H}\mathbf{Q}_{s}(\mathbf{\bar{h}}_{s} \!+\! \mathbf{e}_{s})}{(\mathbf{\bar{h}}_{s} \!+\! \mathbf{e}_{s})^{H}\mathbf{W}(\mathbf{\bar{h}}_{s} \!+\! \mathbf{e}_{s}) \!+\! \sigma_{s}^{2}}, \nonumber\\
 s.t.&&\!\!\!\!\!\!\!\! (\mathbf{\bar{h}}_{e,k} \!+\! \mathbf{e}_{e,k})^{H}\bigg[ \mathbf{Q}_{s} \!-\! \bigg(\frac{1}{t} \!-\! 1\bigg)\mathbf{W} \bigg](\mathbf{\bar{h}}_{e,k} \!+\! \mathbf{e}_{e,k}) \nonumber\\ 
 && \!\leq\! (\frac{1}{t} \!-\! 1)\sigma_{e}^{2},\nonumber\\
&&\!\!\!\!\!\!\!\! \textrm{Tr}(\mathbf{Q}_{s} + \mathbf{W}) \!\leq\! P, ~ \max_{\mathbf{\Delta}_{i}}\textrm{Tr}[(\mathbf{\bar{A}}_{i} \!+\! \mathbf{\Delta}_{i})(\mathbf{Q}_{s} \!+\! \mathbf{W})] \!\leq\! p_{i}, \nonumber\\
&&\!\!\!\!\!\!\!\!\!\!\!\!\!\!\!\!\!\!\!\!\!\!\! (\mathbf{\bar{h}}_{l} \!+\! \mathbf{e}_{l})^{H}(\mathbf{Q}_{s} \!+\! \mathbf{W})(\mathbf{\bar{h}}_{l} \!+\! \mathbf{e}_{l}) \!\geq\! E_{l},~\forall l, \mathbf{Q}_{s} \!\succeq\! \mathbf{0}, \mathbf{W} \!\succeq\! \mathbf{0}, t \!\geq\! 0. \nonumber\\
\end{eqnarray}}
The above problem is not convex due to channel uncertainties. Thus, we consider \emph{S-Procedure} to solve this robust secrecy rate maximization problem, which can be written as  
{\small  \begin{subequations}\label{eq:Robust_sec_rate_relaxed_01}
\begin{eqnarray}
&\!\!\!\!&\!\!\!\! f(t) \!=\! \max_{\mathbf{Q}_{s},\mathbf{W},t,\lambda_{e,k},\alpha_{l}} ~ \frac{(\mathbf{\bar{h}}_{s} + \mathbf{e}_{s})^{H}\mathbf{Q}_{s}(\mathbf{\bar{h}}_{s} + \mathbf{e}_{s})}{(\mathbf{\bar{h}}_{s} + \mathbf{e}_{s})^{H}\mathbf{W}(\mathbf{\bar{h}}_{s} + \mathbf{e}_{s}) + \sigma_{s}^{2}}, \nonumber\\
&&\!\!\!\!\!\!\!\!\!\! s.t.~  \textrm{Tr}(\mathbf{Q}_{s} \!+\! \mathbf{W}) \!\leq\! P, ~ \textrm{Tr}[\mathbf{\bar{A}}_{i}(\mathbf{Q}_{s} \!+\! \mathbf{W})] \!+\! \varepsilon_{i}\|\mathbf{Q}_{s} \!+\! \mathbf{W}\|_{F} \!\leq\! p_{i},\nonumber\\ && ~~~~~\forall i, \label{eq:power_constraint}\!\!\!\!\!\!\!\!\!\!\!\!\!\!\\
&&\!\!\!\!\!\!\!\!\!\!\!\! \left[\!\!\begin{array}{cc}
 \lambda_{e,k}\mathbf{I} \!-\! [\mathbf{Q}_{s} \!-\! (\frac{1}{t} \!-\! 1)\mathbf{W}] \!&\! -[\mathbf{Q}_{s} \!-\! (\frac{1}{t} \!-\! 1)\mathbf{W}]\mathbf{\bar{h}}_{e,k} \\
 -\mathbf{\bar{h}}_{e,k}^{H}[\mathbf{Q}_{s} \!-\! (\frac{1}{t} \!-\! 1)\mathbf{W}] \!&\! c_{k} 
\end{array}
 \!\!\right] \!\succeq\! \mathbf{0}, ~\forall k, \label{eq:Eve_LMI}\nonumber\\\\
&&\!\!\!\!\!\!\!\!\!\!\!\! \left[\!\!\begin{array}{cc}
\alpha_{l}\mathbf{I} \!+\! (\mathbf{Q}_{s} \!+\! \mathbf{W}) \!\!&\!\! (\mathbf{Q}_{s} \!+\! \mathbf{W})\mathbf{\bar{h}}_{l}  \\
\mathbf{\bar{h}}_{l}^{H}(\mathbf{Q}_{s} \!+\! \mathbf{W}) \!\!&\!\! \mathbf{\bar{h}}_{l}^{H}(\mathbf{Q}_{s} \!+\! \mathbf{W})\mathbf{\bar{h}}_{l} \!-\! E_{l} \!-\! \alpha_{l}\varepsilon_{l}^{2}
\end{array}
\!\!\right] \!\succeq\! \mathbf{0}, ~\forall l, \label{eq:EH_LMI}
\end{eqnarray}
\end{subequations}}
where $ c_{k} =  -\mathbf{\bar{h}}_{e,k}^{H}[(\mathbf{Q}_{s} \!-\! \frac{1}{t} \!-\! 1)\mathbf{W}]\mathbf{\bar{h}}_{e,k} \!+\! (\frac{1}{t} \!-\! 1)\sigma_{e}^{2} \!-\! \lambda_{e,k}\varepsilon_{e}^{2} $.
Let us introduce a slack variable $ \tau $ to relax the objective function of the problem \eqref{eq:Robust_sec_rate_relaxed_01}, and by exploiting \emph{S-Procedure} and Charnes-Cooper transformation, we have 
{\small \begin{eqnarray}\label{eq:Sec_rate_max_with_AN_SDP_results}
&&\!\!\!\!\!\!\!\!\!\!\!\! f(t) \!=\! \max_{\mathbf{\bar{Q}}_{s},\mathbf{\bar{W}},\lambda_{s},\mu_{s},\lambda_{e,k},\alpha_{l}} ~ \tau, \nonumber\\
s.t. &&\!\!\!\!\!\!\!\! \left[\begin{array}{cc}
\lambda_{s}\mathbf{I} \!+\! \mathbf{\bar{Q}}_{s} \!&\! \mathbf{\bar{Q}}_{s}\mathbf{\bar{h}}_{s} \\
\mathbf{\bar{h}}_{s}^{H}\mathbf{\bar{Q}}_{s} \!&\! \mathbf{\bar{h}}_{s}^{H}\mathbf{\bar{Q}}_{s}\mathbf{\bar{h}}_{s} \!-\! \tau \!-\! \lambda_{s}\varepsilon_{s}^{2}
\end{array}
\right] \!\succeq\! \mathbf{0}, \nonumber\\
&&\!\!\!\!\!\!\!\! \left[\begin{array}{cc}
\mu_{s}\mathbf{I} \!-\! \mathbf{\bar{W}} \!&\! -\mathbf{\bar{W}}\mathbf{\bar{h}}_{s} \\
-\mathbf{\bar{h}}_{s}^{H}\mathbf{\bar{W}} \!&\! -\mathbf{\bar{h}}_{s}^{H}\mathbf{\bar{W}}\mathbf{\bar{h}}_{s} \!-\! \delta \sigma_{s}^{2} \!+\! 1 \!-\! \mu_{s}\varepsilon_{s}^{2}  
\end{array}
  \right] \!\succeq\! \mathbf{0}, \nonumber\\
&&\!\!\!\!\!\!\!\!\!\!\!\!\!\!\!\! \left[\begin{array}{cc}
\lambda_{e,k}\mathbf{I} \!-\! [\mathbf{\bar{Q}}_{s} \!-\! (\frac{1}{t} \!-\! 1)\mathbf{\bar{W}}] \!&\! -[\mathbf{\bar{Q}}_{s} \!-\! (\frac{1}{t} \!-\! 1)\mathbf{\bar{W}}]\mathbf{\bar{h}}_{e,k}\\
-\mathbf{\bar{h}}_{e,k}^{H}[\mathbf{\bar{Q}}_{s} \!-\! (\frac{1}{t} \!-\! 1)\mathbf{\bar{W}}] \!&\! \bar{c}_{k}
\end{array}
\right] \succeq \mathbf{0}, \nonumber\\
&&\!\!\!\!\!\!\!\!\!\!\!\!\!\!\!\!  \left[\begin{array}{cc}
\alpha_{l}\mathbf{I} \!+\! (\mathbf{\bar{Q}}_{s} \!+\! \mathbf{\bar{W}}) \!&\! (\mathbf{\bar{Q}}_{s} \!+\! \mathbf{\bar{W}})\mathbf{\bar{h}}_{l} \\
 \mathbf{\bar{h}}_{l}^{H}(\mathbf{\bar{Q}}_{s} \!+\! \mathbf{\bar{W}}) \!&\! \mathbf{\bar{h}}_{l}^{H}(\mathbf{\bar{Q}}_{s} \!+\! \mathbf{\bar{W}})\mathbf{\bar{h}}_{l} \!-\! \delta E_{l} \!-\! \alpha_{l}\varepsilon_{l}^{2}
\end{array}
\right] \!\succeq\! \mathbf{0}, \nonumber\\
&& \textrm{Tr}[\mathbf{\bar{A}}_{i}(\mathbf{\bar{Q}}_{s} \!+\! \mathbf{\bar{W}})] \!+\! \varepsilon_{i}\| \mathbf{\bar{Q}}_{s} \!+\! \mathbf{\bar{W}} \|_{F} \!\leq\! \delta p_{i}, \forall i, \nonumber\\
&& \textrm{Tr}(\mathbf{\bar{Q}}_{s} + \mathbf{\bar{W}}) \leq \delta P,
\end{eqnarray} }
where $ \bar{c}_{k} = -\mathbf{\bar{h}}_{e,k}^{H}[\mathbf{\bar{Q}}_{s} \!-\! (\frac{1}{t} \!-\! 1)\mathbf{\bar{W}}]\mathbf{\bar{h}}_{e,k} \!+\! \delta (\frac{1}{t} \!-\! 1) \sigma_{e}^{2} \!-\! \lambda_{e,k}\varepsilon_{e}^{2} $. By solving the problem \eqref{eq:Sec_rate_max_with_AN_SDP_results}, we can obtain the optimal value $ f(t)^{*} $, which can be written based on channel uncertainties as 
{\small  \begin{eqnarray}
&& \frac{(\mathbf{\bar{h}}_{s} + \mathbf{e}_{s})^{H}\mathbf{Q}_{s}(\mathbf{\bar{h}}_{s} + \mathbf{e}_{s})}{(\mathbf{\bar{h}}_{s} + \mathbf{e}_{s})^{H}\mathbf{W}(\mathbf{\bar{h}}_{s} + \mathbf{e}_{s}) + \sigma_{s}^{2}} \geq f(t)^{*}, \nonumber\\
&& \Rightarrow (\mathbf{\bar{h}}_{s} + \mathbf{e}_{s})^{H}[\mathbf{Q}_{s} - f(t)^{*}\mathbf{W}](\mathbf{\bar{h}}_{s} + \mathbf{e}_{s}) \geq f(t)^{*}\sigma_{s}^{2},
\end{eqnarray} }
Thus, the associated power minimization problem can be given as follows:
{\small \begin{subequations}\label{eq:Robust_power_min_results}
\begin{eqnarray}
&\!\!\!\!&\!\!\!\! \min_{\mathbf{Q}_{s},\mathbf{W},\alpha_{l},\beta_{s},\lambda_{e,k}} ~ \textrm{Tr}(\mathbf{Q}_{s}), \nonumber\\
s.t. && \eqref{eq:power_constraint}-\eqref{eq:EH_LMI}, \label{eq:Robust_power_min_constraint01}\\
&&\!\!\!\!\!\!\!\!\!\!\!\!\!\!\!\!\!\! \left[\!\!\begin{array}{cc}
\beta_{s}\mathbf{I} \!+\! [\mathbf{Q}_{s} \!-\! f(t)^{*}\mathbf{W}] \!\!&\!\! [\mathbf{Q}_{s} \!-\! f(t)^{*}\mathbf{W}]\mathbf{\bar{h}}_{s} \\
 \mathbf{\bar{h}}_{s}^{H}[\mathbf{Q}_{s} \!-\! f(t)^{*}\mathbf{W}] \!\!&\!\! d_{s}
\!\!\end{array} 
 \right] \succeq \mathbf{0}, \label{eq:Robust_power_min_constraint02}
\end{eqnarray} 
\end{subequations} }
where $ d_{s} = \mathbf{\bar{h}}_{s}^{H}[\mathbf{Q}_{s} \!-\! f(t)^{*}\mathbf{W}]\mathbf{\bar{h}}_{s} \!-\! f(t)^{*}\sigma_{s}^{2} \!-\! \beta_{s}\varepsilon_{s}^{2} $.
It easily verified that the feasible solution to problem \eqref{eq:Robust_power_min_results} is optimal for \eqref{eq:Robust_sec_rate_relaxed_01}, which is derived from \eqref{eq:Robust_power_min_constraint01} and \eqref{eq:Robust_power_min_constraint02}. Thus, the following theorem holds to show the optimal solution of \eqref{eq:Robust_sec_rate_relaxed_01} is rank-one: 
\begin{theorem}\label{lemma:rank-one_of_robust_sec_rate_max}
Provided that the problem \eqref{eq:Robust_sec_rate_relaxed_01} is feasible, the optimal solution of \eqref{eq:Robust_sec_rate_relaxed_01} always return rank-one, and this optimal solution can be obtained by solving \eqref{eq:Sec_rate_max_with_AN_SDP_results}.
\end{theorem}
\begin{IEEEproof}
Please refer to Appendix \ref{appendix:rank-one_of_robust_sec_rate_max}.
\end{IEEEproof}
\subsubsection{Robust SDP Based Successive Convex Optimization}
Now, we consider the second reformulation for the secrecy rate maximization problem to joint optimization for transmit beamforming and AN covariance matrix, the optimization framework can also be reformulated into robust SDP based SCA by incorporating with channel uncertainty. Thus, this robust secrecy rate maximization problem can be rewritten as  
{\small  \begin{subequations}\label{eq:SCA_robust_sec_rate_max_ori01}
\begin{eqnarray}
 \min_{\mathbf{Q}_{s},\mathbf{W}} && \max_{k} ~\frac{t_{e,k}r_{s}}{t_{s}} \label{eq:SCA_robust_sec_rate_max_channel_uncertainty}\\
s.t. &&\!\!\!\!\!\!\!\!\!\!  \textrm{Tr}(\mathbf{Q}_{s} \!+\! \mathbf{W}) \!\leq\! P, 
~~ \textrm{Tr}[(\mathbf{\bar{A}}_{i}\!+\!\mathbf{\Delta}_{i})(\mathbf{Q}_{s} \!+\! \mathbf{W})] \!\leq\! p_{i},  ~\forall i,
 \label{eq:SCA_robust_power_constraints}\nonumber\\\\
&&\!\!\!\!\!\!\!\! \mathbf{h}_{l}^{H}(\mathbf{Q}_{s} + \mathbf{W})\mathbf{h}_{l} \geq E_{l}, ~\forall l.  \label{eq:SCA_robust_EH_constraint}
\end{eqnarray} 
\end{subequations} }
where $ t_{e,k} = \sigma_{e}^{2} \!+\! (\mathbf{\bar{h}}_{e,k}\!+\!\mathbf{e}_{e,k})^{H}(\mathbf{Q}_{s} \!+\! \mathbf{W})(\mathbf{\bar{h}}_{e,k}\!+\!\mathbf{e}_{e,k}) $, $ r_{s} = \sigma_{s}^{2} \!+\! (\mathbf{\bar{h}}_{s}\!+\!\mathbf{e}_{s})^{H}\mathbf{W}(\mathbf{\bar{h}}_{s}\!+\!\mathbf{e}_{s}) $, $ t_{s} = \sigma_{s}^{2} \!+\! (\mathbf{\bar{h}}_{s}\!+\!\mathbf{e}_{s})^{H}(\mathbf{Q}_{s} \!+\! \mathbf{W})(\mathbf{\bar{h}}_{s}\!+\!\mathbf{e}_{s}) $ and $ r_{e,k} = \sigma_{e}^{2} \!+\! (\mathbf{\bar{h}}_{e,k} \!+\! \mathbf{e}_{e,k})^{H}\mathbf{W}(\mathbf{\bar{h}}_{e,k} \!+\! \mathbf{e}_{e,k}) $.
Let us introduce the following relations for \eqref{eq:SCA_robust_sec_rate_max_channel_uncertainty}
{\small  \begin{subequations}
\begin{eqnarray}
e^{x_{0}} &\!\!\!\!\leq&\!\!\!\! \sigma_{s}^{2} + \min_{\mathbf{e}_{s}}(\mathbf{\bar{h}}_{s}+\mathbf{e}_{s})^{H}(\mathbf{Q}_{s} + \mathbf{W})(\mathbf{\bar{h}}_{s}+\mathbf{e}_{s}),\label{eq:SCA_expoential_variables_channel_uncertainty01}\\ 
e^{x_{k}} &\!\!\!\!\leq&\!\!\!\! \sigma_{e}^{2} + \min_{\mathbf{e}_{e,k}}(\mathbf{\bar{h}}_{e,k}+\mathbf{e}_{e,k})^{H}\mathbf{W}(\mathbf{\bar{h}}_{e,k}+\mathbf{e}_{e,k}), \label{eq:SCA_expoential_variables_channel_uncertainty02}\\
e^{y_{k}} &\!\!\!\!\geq&\!\!\!\! \sigma_{e}^{2} + \max_{\mathbf{e}_{e,k}}(\mathbf{\bar{h}}_{e,k}+\mathbf{e}_{e,k})^{H}(\mathbf{Q}_{s} + \mathbf{W})(\mathbf{\bar{h}}_{e,k}+\mathbf{e}_{e,k}),\label{eq:SCA_expoential_variables_channel_uncertainty03}\\
 e^{y_{0}} &\!\!\!\!\geq&\!\!\!\! \sigma_{s}^{2} + \max_{\mathbf{e}_{s}}(\mathbf{\bar{h}}_{s}+\mathbf{e}_{s})^{H}\mathbf{W}(\mathbf{\bar{h}}_{s}+\mathbf{e}_{s}),\label{eq:SCA_expoential_variables_channel_uncertainty04}
\end{eqnarray}
\end{subequations} }
By employing the slack variables (i.e., $ \tau $, $  u_{s} $, $ u_{e,k} $, $ v_{s} $, and $ v_{e,k} $) for \eqref{eq:SCA_robust_sec_rate_max_channel_uncertainty}, \eqref{eq:SCA_expoential_variables_channel_uncertainty01}-\eqref{eq:SCA_expoential_variables_channel_uncertainty04}, respectively, the problem \eqref{eq:SCA_robust_sec_rate_max_ori01} can be equivalently modified as 
{\small \begin{subequations}
\begin{eqnarray}
&&\!\!\!\!\!\!\!\! \min_{\Omega} ~ \tau, \nonumber\\
s.t. &&\!\!\!\!\!\!\!\! e^{y_{0}\!+\!y_{k}\!-\!x_{0}\!-\!x_{k}} \!\leq\! \tau, ~\forall k,~\eqref{eq:SCA_robust_power_constraints},~\eqref{eq:SCA_robust_EH_constraint},\nonumber\\
&&\!\!\!\!\!\!\!\!\!\!\!\!\!\!\!\!\!\!\!\! e^{x_{0}} \!\leq\! \sigma_{s}^{2} \!+\! u_{s},~\min_{\mathbf{e}_{s}} (\mathbf{\bar{h}}_{s}\!+\!\mathbf{e}_{s})^{H}[\mathbf{Q}_{s}\!+\!\mathbf{W}](\mathbf{\bar{h}}_{s}\!+\!\mathbf{e}_{s}) \!\geq\! u_{s}, \\
&&\!\!\!\!\!\!\!\!\!\!\!\!\!\!\!\!\!\!\!\!\! e^{x_{k}} \!\leq\! \sigma_{e}^{2} \!+\! u_{e,k},~\min_{\mathbf{e}_{e,k}} (\mathbf{\bar{h}}_{e,k}\!+\!\mathbf{e}_{e,k})^{H}\mathbf{W}(\mathbf{\bar{h}}_{e,k}\!+\!\mathbf{e}_{e,k}) \!\geq\! u_{e,k},~\forall k,\nonumber\\\\
&&\!\!\!\!\!\!\!\!\!\!\!\!\!\!\!\!\!\!\!\!\! e^{y_{k}} \!\geq\! \sigma_{e}^{2} \!+\! v_{e,k},~\max_{\mathbf{e}_{e,k}}(\mathbf{\bar{h}}_{e,k}\!+\!\mathbf{e}_{e,k})^{H}(\mathbf{Q}_{s}\!+\!\mathbf{W})(\mathbf{\bar{h}}_{e,k}\!+\!\mathbf{e}_{e,k}) \!\leq\! v_{e,k},\!\!\!\!\!\!\!\!\!\!\!\!\!\!\!\!\!\!\!\!\!\nonumber\\\\
&&\!\!\!\!\!\!\!\! e^{y_{0}} \!\geq\! \sigma_{s}^{2} \!+\! v_{s},~\max_{\mathbf{e}_{s}}(\mathbf{\bar{h}}_{s}\!+\!\mathbf{e}_{s})^{H}\mathbf{W}(\mathbf{\bar{h}}_{s}\!+\!\mathbf{e}_{s}) \!\leq\! v_{s},\\
&&\!\!\!\!\!\!\!\!\!\!\!\!\!\!\!\!\!\!\!\!\! ~
\{\mathbf{Q}_{s},\mathbf{W},\mathbf{e}_{s},\mathbf{e}_{e,k},x_{0},y_{0},x_{k},y_{k},u_{s},u_{e,k},v_{s},v_{e,k}\} \in \Omega.
\end{eqnarray}
\end{subequations} }

By exploiting \emph{S-Procedure} and first-order Taylor series approximation, we have 
{\small \begin{subequations}
\begin{eqnarray}
&&\!\!\!\!\!\!\!\! \min_{\Omega} ~ \tau, \nonumber\\
&&\!\!\!\!\!\!\!\!\!\!\!\!\!\!\! s.t. ~ e^{y_{0}\!+\!y_{k}\!-\!x_{0}\!-\!x_{k}} \!\leq\! \tau, ~ e^{x_{0}} \!\leq\! \sigma_{s}^{2} \!+\! u_{s},~e^{x_{k}} \!\leq\! \sigma_{e}^{2} \!+\! u_{e,k},\\
&&\!\!\!\!\!\!\!\! e^{\bar{y}_{k}}(y_{k} \!-\! \bar{y}_{k} \!+\! 1) \!\geq\! \sigma_{e}^{2} \!+\! v_{e,k},~e^{\bar{y}_{0}}(y_{0} \!-\! \bar{y}_{0} \!+\! 1) \!\geq\! \sigma_{s}^{2} \!+\! v_{s}, \\
&&\!\!\!\!\!\!\!\!\!\!\!\!\!\!\! \left[\begin{array}{cc}
\lambda_{s}\mathbf{I} \!+\! (\mathbf{Q}_{s}\!+\!\mathbf{W})  \!&\! (\mathbf{Q}_{s}\!+\!\mathbf{W})\mathbf{\bar{h}}_{s} \\
\mathbf{\bar{h}}_{s}^{H}(\mathbf{Q}_{s}\!+\!\mathbf{W}) \!&\! \mathbf{\bar{h}}_{s}^{H}(\mathbf{Q}_{s}\!+\!\mathbf{W})\mathbf{\bar{h}}_{s} \!-\! u_{s} \!-\! \lambda_{s}\varepsilon_{s}^{2}
\end{array}
\right] \!\succeq\! \mathbf{0},\\
&&\!\!\!\!\!\!\!\!\!\!\!\!\!\!\! \left[\begin{array}{cc}
\lambda_{e,k}\mathbf{I} \!+\! \mathbf{W}  \!&\! \mathbf{W}\mathbf{\bar{h}}_{e,k} \\
\mathbf{\bar{h}}_{e,k}^{H}\mathbf{W} \!&\! \mathbf{\bar{h}}_{e,k}^{H}\mathbf{W}\mathbf{\bar{h}}_{e,k} \!-\! u_{e,k} \!-\! \lambda_{e,k}\varepsilon_{e,k}^{2}
\end{array}
\right] \succeq \mathbf{0},\\
&&\!\!\!\!\!\!\!\!\!\!\!\!\!\!\! \left[\!\!\begin{array}{cc}
\beta_{e,k}\mathbf{I} \!-\! (\mathbf{Q}_{s}+\mathbf{W})  \!&\! -(\mathbf{Q}_{s}\!+\!\mathbf{W})\mathbf{\bar{h}}_{e,k} \\
-\mathbf{\bar{h}}_{e,k}^{H}(\mathbf{Q}_{s}\!+\!\mathbf{W}) \!&\! -\mathbf{\bar{h}}_{e,k}^{H}(\mathbf{Q}_{s}\!+\!\mathbf{W})\mathbf{\bar{h}}_{e,k} \!+\! v_{e,k} \!-\! \beta_{e,k}\varepsilon_{e,k}^{2}
\end{array}
\!\!\right] \!\succeq\! \mathbf{0},\nonumber\\\\
&&\!\!\!\!\!\!\!\!\!\!\!\!\!\!\! \left[\begin{array}{cc}
\beta_{s}\mathbf{I} \!-\! \mathbf{W}  \!&\! -\mathbf{W}\mathbf{\bar{h}}_{s} \\
-\mathbf{\bar{h}}_{s}^{H}\mathbf{W} \!&\! -\mathbf{\bar{h}}_{s}^{H}\mathbf{W}\mathbf{\bar{h}}_{s} \!+\! v_{s} \!-\! \beta_{s}\varepsilon_{s}^{2}
\end{array}
\right] \!\succeq\! \mathbf{0},\\
&&\!\!\!\!\!\!\!\!\!\!\!\!\!\!\! \left[\begin{array}{cc}
\alpha_{l}\mathbf{I} \!+\! (\mathbf{Q}_{s}\!+\!\mathbf{W})  \!&\! (\mathbf{Q}_{s}\!+\!\mathbf{W})\mathbf{\bar{h}}_{l} \\
\mathbf{\bar{h}}_{l}^{H}(\mathbf{Q}_{s}\!+\!\mathbf{W}) \!&\! \mathbf{\bar{h}}_{l}^{H}(\mathbf{Q}_{s}\!+\!\mathbf{W})\mathbf{\bar{h}}_{l} \!-\! E_{l} \!-\! \alpha_{l}\varepsilon_{l}^{2}
\end{array}
\right] \!\succeq\! \mathbf{0},\\
&&\!\!\!\!\!\!\!\!\!\!\!\!\!\!\! \textrm{Tr}(\mathbf{Q}_{s} \!+\! \mathbf{W}) \!\leq\! P,~ \textrm{Tr}[\mathbf{\bar{A}}_{i}(\mathbf{Q}_{s}\!+\!\mathbf{W})] \!+\! \|\mathbf{Q}_{s}\!+\!\mathbf{W}\|_{F} \!\leq\! p_{i}, \nonumber\\
&&\!\!\!\!\!\!\!\!\!\!\!\!\!\!\! [\mathbf{Q}_{s}\!\succeq\! \mathbf{0},\mathbf{W}\!\succeq\! \mathbf{0},x_{0},y_{0},x_{k},y_{k},u_{s},u_{e,k},v_{s},v_{e,k},\nonumber\\ &&\!\!\!\!\!\!  \lambda_{s}\geq 0,\lambda_{e,k}\geq 0, \beta_{s}\geq 0,\beta_{e,k}\geq 0,\alpha_{l}\geq 0] \!\in\! \Omega,~ \forall i,l,k.
\end{eqnarray}
\end{subequations} }
The above problem is convex for a given $ \bar{y}_{k} $ and $ \bar{y}_{0} $ at each iteration, can be solved by using interior-point method to update the solution for the next iteration until the algorithm converges. Thus the robust SDP based SCA algorithm is similar to Table \ref{Table:SCA_algorithm_robust_sec_rate_max_with_AN_first_new}.
On the other hand, the rank-1 solution can be obtained that is similar to that of one-dimensional line search algorithm shown in \emph{Theorem} \ref{lemma:rank-one_of_robust_sec_rate_max}.

\subsection{Extension to The Case of Multiantenna Eavesdroppers and EH Receivers} \label{subsection_multiple_antennas_case_for_Eves_and_EH_Rx}
Now, we show that the \emph{Theorem} \ref{lemma:rank-one_proof_of_sec_rate_max_with_AN} and \emph{Theorem} \ref{lemma:rank-one_of_robust_sec_rate_max} can also be applied in a more challenging scenario that the eavesdroppers and EH receiver are equipped with multiple antennas by the following \emph{corollary}:
\begin{corollary}
If the eavesdroppers and EH receivers are equipped with multiple antennas, the optimal solution of the problems \eqref{eq:Inner_sec_rate_max_quasi_convex_problem} and \eqref{eq:Robust_sec_rate_relaxed_01} still always return rank-one based on perfect and imperfect CSI.\\ 
\end{corollary}
\begin{IEEEproof}
We first reconsider the constraints \eqref{eq:Eve_rate_slack_variable} and \eqref{eq:EH_constraint_other} based on the assumption that the eavesdroppers and EH receivers are equipped with multiple antennas.  
{\small \begin{subequations}\label{eq:Multiantenna_constraints}
\begin{eqnarray}
&&\!\!\!\!\!\!\!\!\!\!\!\!\!\!\!\!\!\!\!\!\!\! \log \bigg|\!\mathbf{I} \!+\! \bigg(\sigma_{e}^{2}\mathbf{I} \!+\! \mathbf{H}_{e,k}^{H}\mathbf{W}\mathbf{H}_{e,k}\bigg)^{-1}\!\!\mathbf{H}_{e,k}^{H}\mathbf{Q}_{s}\mathbf{H}_{e,k}\!\bigg| \!\leq\! \log(t^{-1}), \label{eq:Multiantenna_eve_rate_constraint}\\
&&\!\!\!\!\!\!\!\!\!\!\!\!\!\!\!\!\!\!\!\!\!\! \textrm{Tr}\bigg[ \mathbf{H}_{l}^{H}(\mathbf{Q}_{s} \!+\! \mathbf{W})\mathbf{H}_{l} \bigg] \!\geq\! E_{l},\label{eq:Multiantenna_EH_constraint}
\end{eqnarray}
\end{subequations} }
where $ \mathbf{H}_{e,k} \in \mathcal{C}^{N_{T} \times N_{E,k}} $ and $ \mathbf{H}_{l} \in \mathcal{C}^{N_{T} \times N_{l}} $ are the channel matrices between the transmitter and the $ k $-th eavesdropper as well as the $ l $-th EH receiver.
The constraint \eqref{eq:Multiantenna_eve_rate_constraint} can be easily relaxed as 
{\small  \begin{eqnarray}\label{eq:Multiantenna_eve_rate_constraint_relaxed}
(\frac{1}{t} - 1)\sigma_{e}^{2}\mathbf{I} + \mathbf{H}_{e,k}^{H}[(\frac{1}{t} - 1)\mathbf{W} - \mathbf{Q}_{s}]\mathbf{H}_{e,k}\succeq \mathbf{0}.
\end{eqnarray} } 
This proof for the above LMI is omitted due to space limit, readers are referred to \cite{Ma_TSP_J13} for more details. 
Since \eqref{eq:Multiantenna_eve_rate_constraint_relaxed} is obtained based on the assumption that $ \textrm{rank}(\mathbf{Q}_{s}) \leq 1 $, we substitute the constraints \eqref{eq:Multiantenna_EH_constraint} and \eqref{eq:Multiantenna_eve_rate_constraint_relaxed} into the associated constraints \eqref{eq:Eve_rate_slack_variable} and \eqref{eq:EH_constraint_other} in the Lagrange dual function \eqref{eq:Lagrange_dual_function_of_power_min_perfect_CSI}, where it easily verified that the optimal solution of $ \mathbf{Q}_{s} $ always returns rank-one. Hence, the problem \eqref{eq:Inner_sec_rate_max_quasi_convex_problem} still has rank-one solution when the eavesdroppers and the EH receivers consist of multiple antennas.
Secondly, we provide a proof for the optimal solution of \eqref{eq:Robust_sec_rate_relaxed_01} still returns rank-one for multiantenna eavesdroppers and EH receivers. The channel uncertainties can be modelled as 
{\small \begin{eqnarray}
\mathbf{H}_{e,k} \!=\! \mathbf{\bar{H}}_{e,k} \!+\! \mathbf{E}_{e,k}, ~\forall k,~
\mathbf{H}_{l} \!=\! \mathbf{\bar{H}}_{l} \!+\! \mathbf{E}_{l},~\forall l, \nonumber
\end{eqnarray} }
where $ \mathbf{\bar{H}}_{e,k} $ and $ \mathbf{\bar{H}}_{l} $ are estimated channel, and $ \mathbf{E}_{e,k} $ and $ \mathbf{E}_{l} $ are channel error matrices, which follow $ \|\mathbf{E}_{e,k}\|_{F} \leq \varepsilon_{k} $ and $ \|\mathbf{E}_{l}\|_{F} \leq \varepsilon_{l} $. Thus, the constraints \eqref{eq:Multiantenna_constraints} can be relaxed by incorporating with channel uncertainties as
{\small  \begin{eqnarray}\label{eq:Multiantenna_Eves_EH_receivers_constraints_LMI}
 \left[\begin{array}{ccc}
\mathbf{A}_{1} \!&\! \mathbf{A}_{2}  \\ 
\mathbf{A}_{2}^{H} \!&\! \mathbf{A}_{3}
\end{array}
\right] \succeq \mathbf{0}, ~ \left[\begin{array}{cc}
\mathbf{B}_{1} \!&\! \mathbf{B}_{2} \\
\mathbf{B}_{2}^{H} \!&\! \mathbf{B}_{3}
\end{array}
\right] \!\succeq\! \mathbf{0},
\end{eqnarray}  }
where $ \mathbf{A}_{1} = [(t^{-1} \!-\! 1)\sigma_{e}^{2} \!-\! \lambda_{e,k}]\mathbf{I} \!+\! \mathbf{\bar{H}}_{e,k}^{H}[(t^{-1} \!-\! 1)\mathbf{W} \!-\! \mathbf{Q}_{s}]\mathbf{\bar{H}}_{e,k} $, $ \mathbf{A}_{2} = \mathbf{\bar{H}}_{e,k}^{H}((t^{-1} \!-\! 1)\mathbf{W} \!-\! \mathbf{Q}_{s}) $, $ \mathbf{A}_{3} = (t^{-1} \!-\! 1)\mathbf{W} \!-\! \mathbf{Q}_{s} \!+\! \frac{\lambda_{e,k}}{\varepsilon_{k}^{2}}\mathbf{I} $, $ \mathbf{B}_{1} = \alpha_{l}\mathbf{I} \!+\! [\mathbf{I}\!\otimes\!(\mathbf{Q}_{s} \!+\! \mathbf{W})] $, $ \mathbf{B}_{2} = [\mathbf{I}\otimes(\mathbf{Q}_{s} \!+\! \mathbf{W})]\mathbf{\bar{h}}_{l} $, and $ \mathbf{B}_{3} = \mathbf{\bar{h}}_{l}^{H}[\mathbf{I}\!\otimes\!(\mathbf{Q}_{s} \!+\! \mathbf{W})]\mathbf{\bar{h}}_{l} \!-\! E_{l} \!-\! \alpha_{l}\varepsilon_{l}^{2} $.
The first LMI of \eqref{eq:Multiantenna_Eves_EH_receivers_constraints_LMI} is obtained based on \cite[Lemma 5]{Ma_TSP_J13}, whereas the second LMI can be relaxed by exploiting \emph{S-Procedure}. Thus, these two constraints are replaced the corresponding constraints in the dual function \eqref{eq:Lagrange_function_of_robust_power_min} so that rank-one solution still can be obtained for the robust secrecy rate maximization problem \eqref{eq:Robust_sec_rate_relaxed_01}.
\end{IEEEproof}

\section{Numerical Results}\label{section Numerical_results}
In this section, we provide the simulation results to validate our proposed algorithm. We consider the a MISO secrecy system in the presence of three eavesdroppers and two EH receivers. The legitimate transmitter is equipped with four transmit antennas (i.e., $N_{T} = 4$), whereas the others are equipped with single antenna. All of channel coefficients (i.e., $ \mathbf{h}_{s} $, $ \mathbf{h}_{e,k} $, and $ \mathbf{h}_{l} $) have been generated as zero-mean circularly symmetric independent and identically distributed Gaussian random variables. The noise power are assumed to be 1 (i.e., $ \sigma_{s}^{2} = \sigma_{e}^{2} = 1 $).
The transmit power (i.e., $ P $) is assumed to be 20dB and all of the error bounds (i.e., $ \varepsilon_{s} $, $ \varepsilon_{e} $ and $ \varepsilon_{l} $) are replaced by $ \varepsilon $ for convenience, which is set to be $ 0.05 $ unless specified.
\subsection{Secrecy Rate Optimization without AN} 
\begin{figure}[!htbp]
\centering
\includegraphics[scale = 0.6]{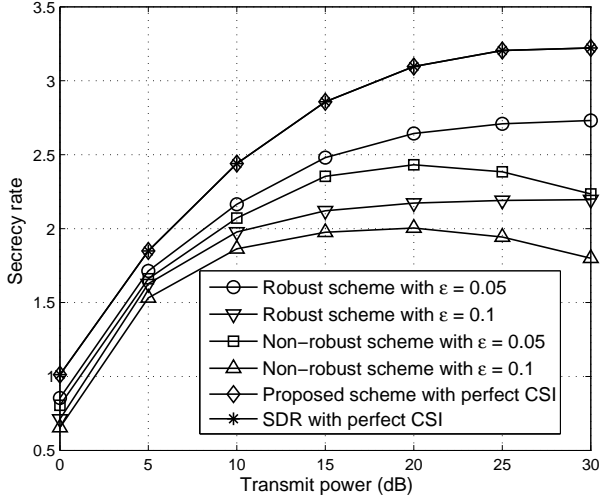}
\caption{Secrecy rate with different transmit powers.}
\label{fig:Robust_sec_rate_with_powers}
\end{figure}

\begin{figure}[!htbp]
\centering
\includegraphics[scale = 0.65]{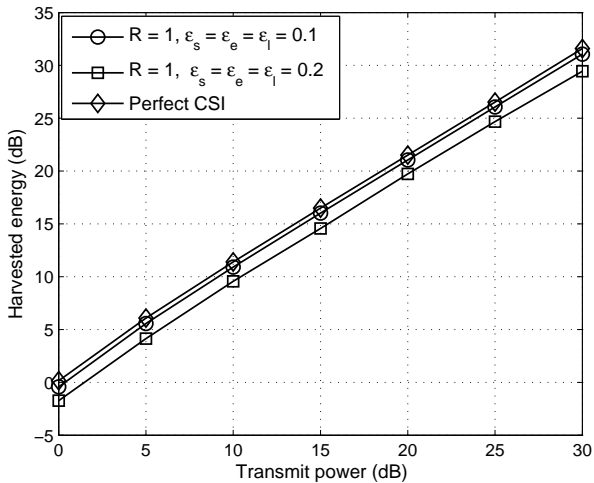}
\caption{Harvested energy with different powers}
\label{fig:Robust_energy_max_energy_with_power}
\end{figure}

\begin{figure}[!htbp]
\centering
\includegraphics[scale = 0.6]{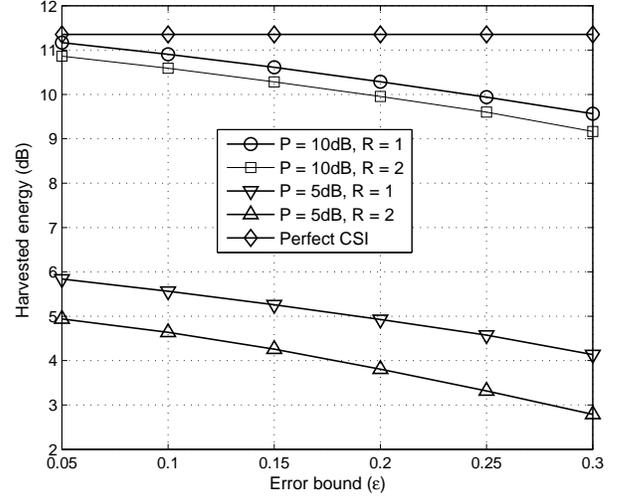}
\caption{Harvested energy with different error bound $ \varepsilon $}
\label{fig:Robust_energy_max_energy_with_error_variances}
\end{figure}

We first provide the simulation results for the secrecy rate maximization problem for perfect and imperfect CSI. Fig. \ref{fig:Robust_sec_rate_with_powers} shows that the achieved secrecy rate with different transmit powers for robust schemes, non-robust schemes and perfect CSI case, where it is easily observed that the achieved secrecy rate increases with transmit power, and our proposed scheme achieves the same performance with the scheme with SDP via rank-relaxation. In addition, the robust scheme outperforms the non-robust scheme in terms of achieved secrecy rate.     
\\
Then, the harvested power with different transmit powers and different error bound $ \varepsilon $ are shown in Fig. \ref{fig:Robust_energy_max_energy_with_power} and Fig. \ref{fig:Robust_energy_max_energy_with_error_variances}.
From these two results, the harvested power increases with the transmit power and decrease with the error bound (i.e., $ \varepsilon $), respectively.  
\subsection{Secrecy Rate Optimization with AN}
\begin{figure}[!htbp]
\centering
\includegraphics[scale = 0.6]{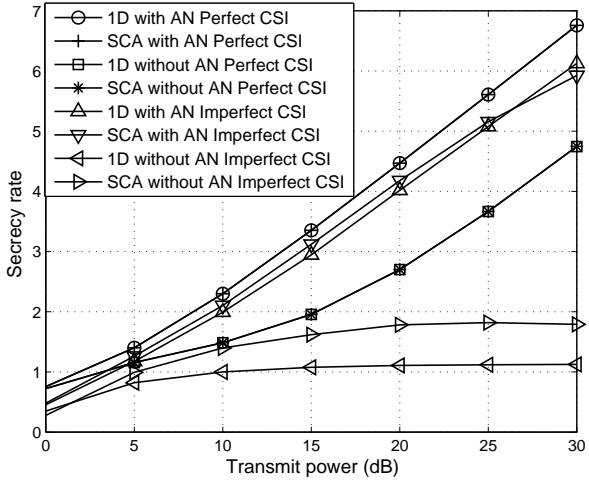}
\caption{The achieved secrecy rate with different transmit power }
\label{fig:Sec_rate_with_AN_vs_powers}
\end{figure}

\begin{figure}[!htbp]
\centering
\includegraphics[scale = 0.6]{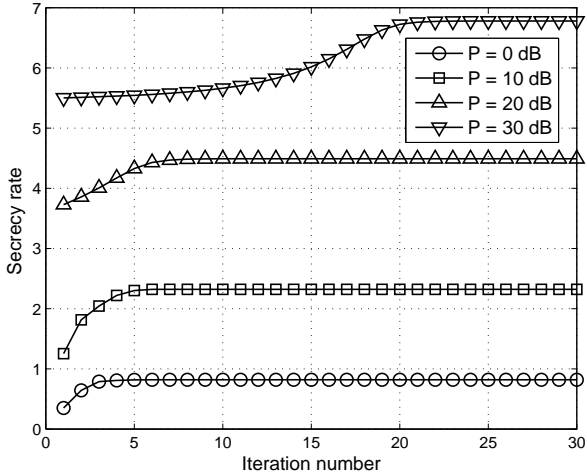}
\caption{The convergence performance of SCA algorithm in different transmit powers}
\label{fig:Convergence_performance_SCA_0_10_20_30}
\end{figure}

\begin{figure}[!htbp]
\centering
\includegraphics[scale = 0.6]{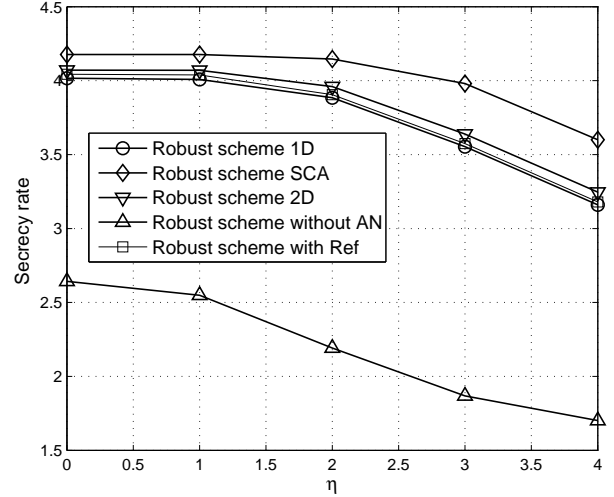}
\caption{Robust secrecy rate with different the ratio of harvested power with transmit power $ \frac{E}{P} $. }
\label{fig:Robust_sec_rate_with_ratio_E_P_new}
\end{figure}

In this subsection, we provide the simulation results to validate the proposed secrecy rate maximization problem for joint optimization of transmit beamforming with AN. Fig. \ref{fig:Sec_rate_with_AN_vs_powers} shows that the achieved secrecy rate with different the ratio of the harvested power with the transmit power (i.e., $ \eta = \frac{E}{P} $) based on perfect and imperfect CSI, where one-dimensional line search based scheme and SCA algorithm with or without AN are plotted, respectively. From this figure, the achieved secrecy rate increase with transmit power, and both schemes in perfect CSI has the same secrecy performance with or without AN, whereas the SCA based scheme has a little better performance than one-dimensional line search based scheme at 0-25  dB, the SCA based scheme outperforms slightly that with one-dimensional line search from 25 to 30 dB. Moreover, the transmit beamforming with AN  has a better performance than that without AN in terms of the achieved secrecy rate. Next, the convergence performance of the SCA based scheme is discussed in Fig. \ref{fig:Convergence_performance_SCA_0_10_20_30}, which shows the convergence of the SCA scheme, from this figure, we can observe that the SCA algorithm converges slower as the transmit power increases. 
Then, the achieved secrecy rate with the ratio of the harvested power $ E $ to the transmit power $ P $ is presented in Fig. \ref{fig:Robust_sec_rate_with_ratio_E_P_new}, where we compare our proposed schemes (i.e., one-dimensional line search and SCA) with two-dimensional search and the scheme shown in \cite{Tian_Maoxin_SPL_J15}. It is observed that one-dimensional line search scheme is closing to the scheme in \cite{Tian_Maoxin_SPL_J15} and two-dimensional search scheme. In addition, the SCA based scheme outperforms these three schemes. At last, we provide the simulation result to show the proportion of the AN power $ \textrm{Tr}(\mathbf{W}) $ in the total transmit power ($ P $) with different signal-to-noise ratio (SNR), which can be replaced by $ P $ due to the noise power equals to 1. One can observe that this proportion increase first and then has a decline after approximately 15 dB, and the robust scheme without the EH receivers has a lower proportion of the AN power than the robust scheme with the EH receivers at the lower SNR regime. In addition, the scheme in \cite{Tian_Maoxin_SPL_J15} has a larger proportion than our both proposed schemes. 

\begin{figure}[!htbp]
\centering
\includegraphics[scale = 0.6]{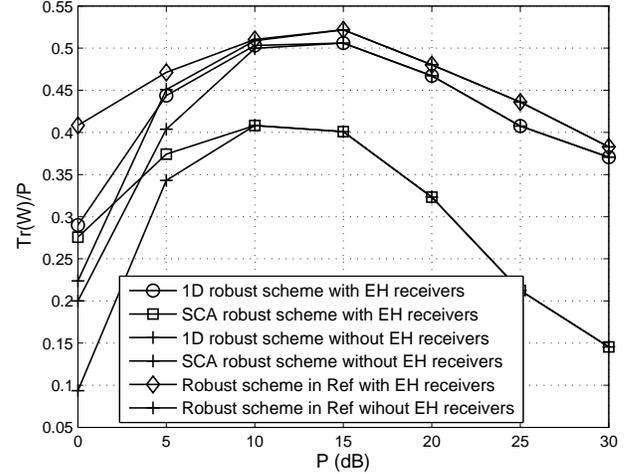}
\caption{The proportion of AN consumption with different SNRs.}
\label{fig:}
\end{figure}
\section{Conclusions}\label{section Conclusions}
In this paper, we investigated the secrecy rate optimization problems for a MISO secrecy channel for SWIPT in presence of multiple eavesdroppers and EH receivers. We first proposed a novel reformulation by using \emph{Nemiroski lemma} to optimize the transmit beamforming to circumvent rank-relaxation based on perfect CSI and imprefect CSI. Then, jointly optimization for transmit beamforming with AN are considered in the secrecy rate maximization problem based on both perfect CSI and imperfect CSI. We reformulated this optimization problem by using one-dimensional line search method and SCA algorithm, respectively. Furthermore, we provide the tightness analysis for this optimization problem, which can show the optimal solution of this secrecy rate maximization problem exactly returns rank-one. Simulation results have been provided to validate the performance of our proposed schemes. 
\begin{appendix}  
\subsection{Proof of Lemma \ref{lemma:Schur_complement_applications}}\label{appendix:Schur_complement_applications}
In order to prove \emph{Lemma 1}, we first rewrite the secrecy rate constraint in \eqref{eq:power_min_modified} as
{\small  \begin{eqnarray}\label{eq:sec_rate_constraint_SOCP}
\frac{1}{\sigma_{s}^{2}}|\mathbf{w}^{H}\mathbf{h}_{s}|^{2} \!\geq\! \left[\!\!\begin{array}{cc}
\frac{2^{\frac{R}{2}}}{\sigma_{e}}\mathbf{w}^{H}\mathbf{h}_{e,k} \\ (2^{R} - 1)^{\frac{1}{2}}
\end{array}\!\!\right]^{H}\!\!
\left[\!\!\begin{array}{cc}
\frac{2^{\frac{R}{2}}}{\sigma_{e}}\mathbf{w}^{H}\mathbf{h}_{e,k} \\ (2^{R} - 1)^{\frac{1}{2}}
\end{array}\!\!\right]
\end{eqnarray} }
Then, the following lemma is required to convert the above constraint into a SDP:\\
\begin{lemma}
(Schur complement) \cite{boyd_B04}: Let $ \mathbf{X} $ is complex hermitian matrix, 
{\small  \begin{eqnarray}
\mathbf{X} \!=\! \mathbf{X}^{H} \!=\! \left[\!\!\begin{array}{cc}
\mathbf{A} \!&\! \mathbf{B} \\
\mathbf{B}^{H} \!&\! \mathbf{C} 
\end{array}
\!\!\right]
\end{eqnarray} }
Thus, $ \mathbf{S} = \mathbf{C} - \mathbf{B}^{H}\mathbf{A}^{-1}\mathbf{B} $ is the \emph{Schur complement} of $ \mathbf{A} $ in $ \mathbf{X} $, and the following statements hold:
\begin{itemize}
\item $ \mathbf{X} \succ \mathbf{0} $, if and only if $ \mathbf{A} \succ \mathbf{0} $ and $ \mathbf{S} \succ \mathbf{0} $.
\item if $ \mathbf{A} \succ \mathbf{0} $ then $ \mathbf{X} \succ \mathbf{0} $ if and only if $ \mathbf{S} \succ \mathbf{0} $.
\end{itemize} 
\end{lemma}
By exploiting the \emph{Schur complement}, \eqref{eq:sec_rate_constraint_SOCP} can be reformulated as 
{\small  \begin{eqnarray}
\left[\!\!\begin{array}{cc}
 \frac{1}{\sigma_{s}}\mathbf{w}^{H}\mathbf{h}_{s}\mathbf{I} \!\!&\!\! \left[\begin{array}{cc}
\frac{2^{\frac{R}{2}}}{\sigma_{e}}\mathbf{w}^{H}\mathbf{h}_{e,k} \\ (2^{R} - 1)^{\frac{1}{2}}
 \end{array}
\right]   \\ 
 \left[\begin{array}{cc}
\frac{2^{\frac{R}{2}}}{\sigma_{e}}\mathbf{w}^{H}\mathbf{h}_{e,k} \\ (2^{R} - 1)^{\frac{1}{2}}
 \end{array}
\right]^{H} \!\!&\!\! \frac{1}{\sigma_{s}}\mathbf{w}^{H}\mathbf{h}_{s}
\end{array} 
 \!\!\right] \succeq \mathbf{0},
\end{eqnarray} }
In addition, we consider the reformulation of the EH constraint in \eqref{eq:power_min_modified}. In order to express this constraint clearly, we introduce two variables (i.e., $ x_{l} \in \mathbb{R} $ and $ y_{l} \in \mathbb{R} $) such that this constraint can be equivalently modified as 
{\small \begin{subequations}
\begin{eqnarray}
&\!\!\!\!&\!\!\!\! x_{l}^{2} + y_{l}^{2} \geq E, \label{eq:EH_modified_01}\\ 
&\!\!\!\!&\!\!\!\! x_{l} = \Re\{ \mathbf{w}^{H}\mathbf{h}_{l} \}, ~ y_{l} = \Im\{ \mathbf{w}^{H}\mathbf{h}_{l} \}, ~ \forall l. \label{eq:EH_modified_02}
\end{eqnarray}
\end{subequations}}
The constraint \eqref{eq:EH_modified_02} is convex (linear), whereas \eqref{eq:EH_modified_01} is not still convex, thus, first-order Taylor approximation is considered to obtain the desired upper bound. \\ 
Setting $ \mathbf{u}_{l} = [x_{l} ~ y_{l}]^{T} $, thus, $ x_{l}^{2} + y_{l}^{2} = \mathbf{u}_{l}^{T}\mathbf{u}_{l} $. $ \mathbf{u}_{l}^{(n)} $ is $ n $-th iteration of the vector $ \mathbf{u}_{l} $. Thus, \eqref{eq:EH_modified_01} can be approximated as 
{\small \begin{eqnarray}
\mathbf{u}_{l}^{T}\mathbf{u}_{l} \approx \|\mathbf{u}_{l}^{(n)}\|^{2} + 2 \sum_{i = 1}^{2} \mathbf{u}_{l}^{(n)}(i) [ \mathbf{u}_{l}(i) - \mathbf{u}_{l}^{(n)}(i) ],
\end{eqnarray} }
where $ i $ denotes the $ i $-th element of the vector $ \mathbf{u}_{l} $.
\\ This completes \emph{Lemma} \ref{lemma:Schur_complement_applications}.~~~~~~~~~~~~~~~~~~~~~~~~~~~~~~~~~~~~~~~~~~~~~~~~$ \blacksquare $\\
\subsection{Proof of Lemma  \ref{lemma:Nemirovski_lemma_application}}\label{appendix:proof_of_Nemirovski_lemma}
The second constraint in \eqref{eq:sec_rate_constraint_SCA_approximation} can be rewritten as 
{\small \begin{eqnarray}\label{eq:LMI_schur_complement}
\!\!\!\!\!\!\!\!\!\! \mathbf{S}^{'}_{k} \!\!=\!\! \left[\!\!\begin{array}{cc}
f^{(n)}(t_{2})\mathbf{I} \! &\! \left[\!\!\begin{array}{cc}
\frac{2^{\frac{R}{2}}}{\sigma_{e}}\mathbf{w}^{H}(\mathbf{\bar{h}}_{e,k}\!+\!\mathbf{e}_{e,k}) \\ (2^{R}\!-\!1)^{\frac{1}{2}}
\end{array}
\!\!\right]  \\ 
\left[\!\!\begin{array}{cc}
\frac{2^{\frac{R}{2}}}{\sigma_{e}}\mathbf{w}^{H}(\mathbf{\bar{h}}_{e,k}\!+\!\mathbf{e}_{e,k}) \\ (2^{R}\!-\!1)^{\frac{1}{2}}
\end{array}
\!\!\right]^{H}  \!&\! f^{(n)}(t_{2})
\end{array}
\!\!\right] \!\succeq\! \mathbf{0},\!\!\!\!\!\!\!\!\!\!\!\!\!\!\!\nonumber\\
\end{eqnarray}}
\begin{lemma}
(\emph{Nemirovski lemma}) \cite{Nemirovski_TSP_J05}: For a given set of matrices $ \mathbf{A} = \mathbf{A}^{H} $, $ \mathbf{B} $ and $ \mathbf{C} $, the following linear matrix inequality is satisfied:
{\small \begin{eqnarray}
\mathbf{A} \succeq \mathbf{B}\mathbf{X}\mathbf{C}+\mathbf{C}^{H}\mathbf{X}^{H}\mathbf{B}, \|\mathbf{X}\|\leq t,
\end{eqnarray} }
if and only if there exist non-negative real numbers $ a $ such that 
{\small \begin{eqnarray}
\left[\begin{array}{cc}
      \mathbf{A} - a\mathbf{C}^{H}\mathbf{C} & - t\mathbf{B}^{H} \\
    - t\mathbf{B}  &  a \mathbf{I}
    \end{array}\right] \succeq 0.
\end{eqnarray} }
\end{lemma}
By exploiting \emph{Nemirovski lemma}, the constraint in \eqref{eq:LMI_schur_complement} is written as
{\small \begin{eqnarray}
\mathbf{S}_{k} \!\succeq\! \left[\!\!\begin{array}{cc}
\frac{2^{\frac{R}{2}}}{\sigma_{e}}\mathbf{w}^{H}  \!\\\! \mathbf{0} \!\\\! \mathbf{0}
\end{array}
\!\!\right] \mathbf{e}_{e,k} \left[\!\! \begin{array}{cc}
 \mathbf{0} \!&\! -1
\end{array}
 \!\!\right] \!+\! \left[\!\!\begin{array}{cc}
  \mathbf{0} \!\\\! -1
 \end{array}
 \!\!\right] \mathbf{e}_{e,k}^{H}\left[\!\!\begin{array}{ccc}
\frac{2^{\frac{R}{2}}}{\sigma_{e}}\mathbf{w}  \!&\! \mathbf{0} \!&\! \mathbf{0}
\end{array}
\!\!\right],
\end{eqnarray} }
where 
{\small \begin{eqnarray}
\mathbf{S}_{k} = \left[\begin{array}{cc}
 f^{(n)}(t_{2})\mathbf{I} & \left[\!\!\begin{array}{cc}
\frac{2^{\frac{R}{2}}}{\sigma_{e}}\mathbf{w}^{H}\mathbf{\bar{h}}_{e,k} \\ (2^{R}\!-\!1)^{\frac{1}{2}}
\end{array}
\!\!\right] \\
\left[\!\!\begin{array}{cc}
\frac{2^{\frac{R}{2}}}{\sigma_{e}}\mathbf{w}^{H}\mathbf{\bar{h}}_{e,k} \\ (2^{R}\!-\!1)^{\frac{1}{2}}
\end{array}
\!\!\right]^{H}  & f^{(n)}(t_{2})
\end{array}
\right]
\end{eqnarray}}
Thus, the robust secrecy rate constraint can be reformulated as 
{\small \begin{eqnarray}
\mathbf{\bar{S}}_{k} \!=\! \left[\!\!\!\begin{array}{cc}
\mathbf{S}_{k} \!-\! \lambda_{k}\left[\!\!\begin{array}{cc}
\mathbf{0} \!&\! -1
\end{array}\!\!\right] \left[\!\!\begin{array}{cc}
\mathbf{0} \\ -1
\end{array}\!\!\right] \!&\! -\varepsilon_{e,k}\left[\!\!\begin{array}{cc}
\frac{2^{\frac{R}{2}}}{\sigma_{e}}\mathbf{w}^{H}  \!\\\! \mathbf{0} \!\\\! \mathbf{0}
\end{array}
\!\!\right] \\ 
 -\varepsilon_{e,k} \left[\!\!\begin{array}{ccc}
\frac{2^{\frac{R}{2}}}{\sigma_{e}}\mathbf{w}^{H}  \!&\! \mathbf{0} \!&\! \mathbf{0}
\end{array}
\!\!\right]   \!&\!  \lambda_{k}\mathbf{I}
\end{array}\!\!\!\right] \!\succeq\! \mathbf{0}, \forall k.
\end{eqnarray}  }
This completes \emph{Lemma} \ref{lemma:Nemirovski_lemma_application}.~~~~~~~~~~~~~~~~~~~~~~~~~~~~~~~~~~~~~~~~~~~~$  \blacksquare $\\
\subsection{Proof of Theorem \ref{lemma:f(tau)}}\label{appendix:f(tau)}
In order to prove \emph{Theorem 1}, it is assumed that $ \varphi_{1} $ and $ \varphi_{2} $ are optimal values of \eqref{eq:Robust_energy_max_ori} and \eqref{eq:Robust_energy_max_equivalent_optimal}, respectively. We first show \eqref{eq:Robust_energy_max_equivalent_optimal} can obtain the optimal value of \eqref{eq:Robust_energy_max_ori} (i.e., $ \varphi_{2} \geq \varphi_{1} $). It easily verified that the following equality holds:
{\small  \begin{eqnarray}
\varphi_{1} = f(\tau^{*}).
\end{eqnarray} }
On the other hand, $ \varphi_{2} = \max_{\tau \geq 0} f(\tau) \geq f(\tau^{*}) $, thus, $ \varphi_{2} \geq \varphi_{1} $. Secondly, in order to show that \eqref{eq:Robust_energy_max_ori} achieve the optimal value of  \eqref{eq:Robust_energy_max_equivalent_optimal} (i.e., $ \varphi_{1} \geq \varphi_{2} $). Here, we assume that $ \tau^{\dagger} $ is the optimal value of \eqref{eq:Robust_energy_max_equivalent_optimal}, and $ \mathbf{w}^{\dagger} $ is the optimal solution of \eqref{eq:Robust_energy_max_relax_ori} with $ \tau = \tau^{\dagger} $. It can be easily observed from \eqref{eq:Robust_sec_rate_constraint_without_AN}, \eqref{eq:Eve_constraint_relax}, and \eqref{eq:User_constraint_relax} that $ \mathbf{w}^{\dagger} $ is a feasible solution of \eqref{eq:Robust_energy_max_ori}, thus, $ \varphi_{1} \geq \varphi_{2} $. We combine these two parts, it includes $ \varphi_{1} = \varphi_{2} $. \\
This completes \emph{Theorem} \ref{lemma:f(tau)}.~~~~~~~~~~~~~~~~~~~~~~~~~~~~~~~~~~~~~~~~~~~$ \blacksquare $
\subsection{Proof of Theorem \ref{lemma:rank-one_proof_of_sec_rate_max_with_AN}}\label{appendix:rank-one_proof_of_sec_rate_max_with_AN}
The Lagrange function of \eqref{eq:power_min_to_rank_one} can be written as 
{\small  \begin{eqnarray}\label{eq:Lagrange_dual_function_of_power_min_perfect_CSI}
&&\!\!\!\!\!\!\!\!\!\!\! \mathcal{L}(\mathbf{Q}_{s},\mathbf{Y},\mathbf{Z},\lambda,\mu,\eta_{i},\nu_{l},\tau_{k}) \!=\! \textrm{Tr}(\mathbf{Q}_{s}) \nonumber\\ 
&&\!\!\!\!\!\!\!\!\! -  \lambda \bigg[ \textrm{Tr}[\mathbf{h}_{s}\mathbf{h}_{s}^{H}(\mathbf{Q}_{s} \!-\! f(t)\mathbf{W})] \!-\! \sigma_{s}^{2}f(t) \bigg]  \!\!+\!\! \mu \bigg[ \textrm{Tr}(\mathbf{Q}_{s} \!+\! \mathbf{W}) \!-\! P \bigg]\nonumber\\
&&\!\!\!\!\!\!\!\!\!
 \!+ \sum_{i=1}^{N_{T}} \eta_{i} \bigg[ \textrm{Tr}[\mathbf{A}_{i}(\mathbf{Q}_{s} \!+\! \mathbf{W})] \!-\! p_{i} \bigg] \!-\! \sum_{l=1}^{L} \nu_{l}\bigg[ \textrm{Tr}[\mathbf{h}_{l}\mathbf{h}_{l}^{H}(\mathbf{Q}_{s} \!+\! \mathbf{W})] \nonumber\\ 
 &&\!\!\!\!\!\!\!\!\! \!-\! E_{l} \bigg]  \!+\! \sum_{k=1}^{K} \tau_{k} \bigg[ \textrm{Tr}[\mathbf{h}_{e,k}\mathbf{h}_{e,k}^{H}(\mathbf{Q}_{s} \!-\! (t\!-\!1)\mathbf{W})] \!-\! (t-1)\sigma_{e}^{2} \bigg] \nonumber\\ 
 &&\!\!\!\!\!\!\!\!\! \!-\! \textrm{Tr}(\mathbf{Y}\mathbf{Q}_{s}) \!-\! \textrm{Tr}(\mathbf{Z}\mathbf{W}),\!\!\!\!\!\!\!\!\!
\end{eqnarray} }
where $ \mathbf{Y} $, $ \mathbf{Z} $, $ \lambda $, $ \mu $, $ \eta_{i} $, $ \nu_{l} $, $ \tau_{k} $ denote the dual variables of $ \mathbf{Q}_{s} $, $ \mathbf{W} $, \eqref{eq:f(t)_constraint}, \eqref{eq:Power_constraints}, \eqref{eq:EH_constraint_other}, and \eqref{eq:Eve_rate_constraint_reformulation}, respectively. 
Then, we consider the following related KKT conditions: 
{\small \begin{subequations}
\begin{eqnarray}
\frac{\partial \mathcal{L}}{\partial \mathbf{Q}_{s}} &\!\!\!\!=&\!\!\!\! 0, \Rightarrow \mathbf{Y} \!=\! \mathbf{I} \!-\! \lambda\mathbf{h}_{s}\mathbf{h}_{s}^{H} \!+\! \mu\mathbf{I} \!+\! \sum_{i=1}^{N_{T}} \eta_{i}\mathbf{A}_{i} \!-\! \sum_{l=1}^{L} \nu_{l}\mathbf{h}_{l}\mathbf{h}_{l}^{H} \nonumber\\ 
&& \!+\! \sum_{k=1}^{K}\tau_{k}\mathbf{h}_{e,k}\mathbf{h}_{e,k}^{H}, \label{eq:KKT_derivatives_with_Qs}\\
\frac{\partial \mathcal{L}}{\partial \mathbf{W}} &\!\!\!\!=&\!\!\!\! 0, \Rightarrow \mathbf{Z} \!=\! \lambda f(t)\mathbf{h}_{s}\mathbf{h}_{s}^{H} \!+\! \mu\mathbf{I} \!+\! \sum_{i=1}^{N_{T}}\eta_{i}\mathbf{A}_{i} \!-\! \sum_{l=1}^{L}\nu_{l}\mathbf{h}_{l}\mathbf{h}_{l}^{H} \nonumber\\ 
&& \!-\! \sum_{k=1}^{K}\tau_{k}(t-1)\mathbf{h}_{e,k}\mathbf{h}_{e,k}^{H}, \label{eq:KKT_derivatives_with_W}\\
\mathbf{Q}_{s}\mathbf{Y} &\!\!\!\!=&\!\!\!\! \mathbf{0}, \mathbf{Z} \succeq \mathbf{0},~\lambda \geq 0,~ \forall i, l, k. \label{eq:KKT_another}
\end{eqnarray}
\end{subequations} }
By subtracting \eqref{eq:KKT_derivatives_with_W} from \eqref{eq:KKT_derivatives_with_Qs}, we have
{\small \begin{eqnarray}\label{eq:modified_KKT_0102}
\mathbf{Y} - \mathbf{Z} = \mathbf{I} - \lambda(1+f(t))\mathbf{h}_{s}\mathbf{h}_{s}^{H} + \sum_{k=1}^{K} \tau_{k} t \mathbf{h}_{e,k}\mathbf{h}_{e,k}^{H},\nonumber\\
\Rightarrow \mathbf{Y} = \mathbf{A} - \lambda(1+f(t))\mathbf{h}_{s}\mathbf{h}_{s}^{H},
\end{eqnarray} }
where $ \mathbf{A} = \mathbf{I} + \mathbf{Z} + \sum_{k=1}^{K} \tau_{k} t \mathbf{h}_{e,k}\mathbf{h}_{e,k}^{H} $. From \eqref{eq:modified_KKT_0102}, one can easily observe that $ \mathbf{A} $ is positive definite, and $ \textrm{rank}(\mathbf{A}) = N_{T} $, whereas $ \textrm{rank}(\mathbf{Y}) = N_{T} $ or $ N_{T} -1 $. However, $ \mathbf{Q}_{s} = \mathbf{0} $ when $ \textrm{rank}(\mathbf{Y}) = N_{T} $ due to the first condition in \eqref{eq:KKT_another}, which violates $ R_{s} - R_{e,k} > 0 $. In addition, it is easily verified that $ \lambda > 0 $ and $ f(t) >0 $. Thus, $ \textrm{rank}(\mathbf{Y}) = N_{T} - 1 $ always holds, which implies $ \mathbf{Q}_{s} $ lies in the null space of $ \mathbf{Y} $ from \eqref{eq:KKT_another}, thus $ \textrm{rank}(\mathbf{Q}_{s}) = 1 $.\\
This completes \emph{Theorem 2}.~~~~~~~~~~~~~~~~~~~~~~~~~~~~~~~~~~~~~~~~~~~~~$ \blacksquare $

\subsection{Proof of Theorem \ref{lemma:rank-one_of_robust_sec_rate_max}}\label{appendix:rank-one_of_robust_sec_rate_max}
We first consider the dual function of the problem \eqref{eq:Robust_power_min_results},  
{\small \begin{eqnarray}\label{eq:Lagrange_function_of_robust_power_min}
&&\!\!\!\!\!\!\!\!\!\!\!\! \mathcal{L}(\mathbf{Q}_{s},\mathbf{W},\mathbf{Y},\mathbf{Z},\lambda , \gamma_{i},\mathbf{T}_{s},\mathbf{T}_{e,k},\mathbf{T}_{l}) \!=\! \textrm{Tr}(\mathbf{Q}_{s}) \!-\! \textrm{Tr}(\mathbf{Y}\mathbf{Q}_{s}) \nonumber\\ 
&&\!\!\!\! \!-\! \textrm{Tr}(\mathbf{Z}\mathbf{W}) + \lambda [ \textrm{Tr}(\mathbf{Q}_{s} \!+\! \mathbf{W}) \!-\! P ]  \!+\! \sum_{i=1}^{N_{T}} \gamma_{i}\bigg[ \textrm{Tr}[\mathbf{\bar{A}}_{i}(\mathbf{Q}_{s} \!+\! \mathbf{W})] \nonumber\\
&&\!\!\!\! \!+ \varepsilon_{i}\|\mathbf{Q}_{s}+\mathbf{W}\|_{F} \!-\! p_{i} \bigg] \!-\! \textrm{Tr}(\mathbf{T}_{s}\mathbf{A}_{1}) \!-\! \textrm{Tr}[\mathbf{T}_{s}\mathbf{H}_{s}^{H}(\mathbf{Q}_{s} \!-\! f(t)\mathbf{W})\mathbf{H}_{s}] \nonumber\\
&&\!\!\!\! \!-\! \sum_{k=1}^{K} \textrm{Tr}(\mathbf{T}_{e,k}\mathbf{B}_{k}) \!+\! \sum_{k=1}^{K} \textrm{Tr}\bigg[\mathbf{T}_{e,k}\mathbf{H}_{e,k}^{H}[\mathbf{Q}_{s} \!-\! (t^{-1} \!-\! 1)\mathbf{W}]\mathbf{H}_{e,k}\bigg]\nonumber\\
&&\!\!\!\!
 \!-\! \sum_{l=1}^{L}\textrm{Tr}(\mathbf{T}_{l}\mathbf{C}_{l})  \!-\! \sum_{l=1}^{L}\textrm{Tr}[\mathbf{T}_{l}\mathbf{H}_{l}^{H}(\mathbf{Q}_{s} \!+\! \mathbf{W})\mathbf{H}_{l}], \!\!\!\!
\!\!\!\!\!\!\!\!
\end{eqnarray}}
where $ \mathbf{Y} \in \mathbb{H}_{+}^{N_{T}} $, $ \mathbf{Z} \in \mathbb{H}_{+}^{N_{T}} $, $ \lambda \in \mathbb{R}_{+} $, $ \gamma_{i} \in \mathbb{R}_{+} $, $ \mathbf{T}_{s} \in \mathbb{H}_{+}^{N_{T}} $, $ \mathbf{T}_{e,k} \in \mathbb{H}_{+}^{N_{T}} $ and $ \mathbf{T}_{l} \in \mathbb{H}_{+}^{N_{T}}  $ are dual variables of $ \mathbf{Q}_{s} $, $ \mathbf{W} $, \eqref{eq:power_constraint}, \eqref{eq:EH_LMI} and \eqref{eq:Eve_LMI}, respectively, in addition, 
{\small \begin{eqnarray}
\mathbf{A}_{1} &\!\!\!\!\!\!=&\!\!\!\!\!\! \left[\begin{array}{cc}
\beta_{s}\mathbf{I} \!&\! \mathbf{0} \\
\mathbf{0}^{H} \!&\! -f(t)\sigma_{s}^{2} \!-\! \beta_{s}\varepsilon_{s}^{2} 
\end{array}
\right], ~\mathbf{H}_{s} \!=\! \left[\begin{array}{cc}
 \mathbf{I}_{N_{T}} \!&\! \mathbf{\bar{h}}_{s}
\end{array}
\right], \nonumber\\
\mathbf{B}_{k} &\!\!\!\!\!\!=&\!\!\!\!\!\! \left[\begin{array}{cc}
\lambda_{e,k}\mathbf{I} \!&\! \mathbf{0} \\
\mathbf{0}^{H} \!&\! (t^{-1} \!-\! 1)\sigma_{e}^{2} \!-\! \lambda_{e,k}\varepsilon_{e}^{2} 
\end{array}
\right],~\mathbf{H}_{e,k} \!=\! \left[\begin{array}{cc}
 \mathbf{I}_{N_{T}} \!&\! \mathbf{\bar{h}}_{e,k}
\end{array}
\right], \nonumber\\
\mathbf{C}_{l} &\!\!\!\!\!\!=&\!\!\!\!\!\! \left[\begin{array}{cc}
\alpha_{l}\mathbf{I} \!&\! \mathbf{0} \\
\mathbf{0}^{H} \!&\! -E_{l} \!-\! \alpha_{l}\varepsilon_{l}^{2} 
\end{array}
\right],~ \mathbf{H}_{l} \!=\! \left[\begin{array}{cc}
 \mathbf{I}_{N_{T}} \!&\! \mathbf{\bar{h}}_{l}
\end{array}
\right]. \nonumber
\end{eqnarray} }

The related KKT conditions are considered as follows:
{\small \begin{subequations}
\begin{eqnarray}
 \frac{\partial \mathcal{L}}{\partial \mathbf{Q}_{s}} &\!\!\!\!\!\!=&\!\!\!\!\!\! 0,\Rightarrow \mathbf{Y}  \!= \! \mathbf{I} \!+\! \lambda\mathbf{I} \!+\! \sum_{i=1}^{N_{T}} \gamma_{i} [\mathbf{\bar{A}}_{i} \!+\! \varepsilon_{i}\|\mathbf{Q}_{s} \!+\! \mathbf{W}\|_{F}^{-1}\mathbf{I}]  \nonumber\\
  &&\!\!\!\!\!\!\!\!\!\!\!\! \!-\! \mathbf{H}_{s}\mathbf{T}_{s}\mathbf{H}_{s}^{H} \!+\! \sum_{k=1}^{K} \mathbf{H}_{e,k}\mathbf{T}_{e,k}\mathbf{H}_{e,k}^{H} \!-\!  \sum_{l=1}^{L} \mathbf{H}_{l}\mathbf{T}_{l}\mathbf{H}_{l}^{H}, \label{eq:KKT_Qs}\\
 \frac{\partial \mathcal{L}}{\partial \mathbf{W}} &\!\!\!\!\!\!=&\!\!\!\!\!\! 0,  \! \Rightarrow \! \mathbf{Z} \!=\! \lambda\mathbf{I} \!+\! \sum_{i=1}^{N_{T}}\! \gamma_{i} [\mathbf{\bar{A}}_{i} \!+\! \varepsilon_{i}\|\mathbf{Q}_{s} \!+\! \mathbf{W}\|_{F}^{-1}\mathbf{I}] \!+\! f(t)\mathbf{H}_{s}\mathbf{T}_{s}\mathbf{H}_{s}^{H} \nonumber\\ 
 &&\!\!\!\!\!\!\!\!\!\!\!\!  \!-\! \sum_{k=1}^{K} (t^{-1} \!-\! 1)\mathbf{H}_{e,k}\mathbf{T}_{e,k}\mathbf{H}_{e,k}^{H} \!-\! \sum_{l=1}^{L} \mathbf{H}_{l}\mathbf{T}_{l}\mathbf{H}_{l}^{H}, \label{eq:KKT_W} \\
\mathbf{Q}_{s}\mathbf{Y} &\!\!\!\!=&\!\!\!\! \mathbf{0}, \mathbf{Z} \!\succeq\! \mathbf{0}, ~\forall i,k,l, \label{eq:KKT_another}\\
&&\!\!\!\!\!\!\!\!\!\!\!\!\!\!\!\!\!\!\!\!\!\!\!\!\!\! [\mathbf{A}_{1}+\mathbf{H}_{s}^{H}(\mathbf{Q}_{s} - f(t)\mathbf{W})\mathbf{H}_{s}]\mathbf{T}_{s} = \mathbf{0} \label{eq:KKT_Main_channel_LMI}.
\end{eqnarray}
\end{subequations} }
By subtracting \eqref{eq:KKT_W} from \eqref{eq:KKT_Qs}, we have
{\small  \begin{eqnarray}\label{eq:Y_substract_Z}
&&\!\!\!\!\!\!\!\!\!\!\!\!\!\!\!\! \mathbf{Y} \!-\! \mathbf{Z} \!=\! \mathbf{I} \!-\! [1 \!+\! f(t)]\mathbf{H}_{s}\mathbf{T}_{s}\mathbf{H}_{s}^{H} \!+\! \sum_{k=1}^{K} t^{-1} \mathbf{H}_{e,k}\mathbf{T}_{e,k}\mathbf{H}_{e,k}^{H},\nonumber\\
&&\!\!\!\!\!\!\!\!\!\!\!\!\!\!\!\! \Rightarrow  \mathbf{Y} \!+\! [1 \!+\! f(t)]\mathbf{H}_{s}\mathbf{T}_{s}\mathbf{H}_{s}^{H} \!=\! \mathbf{I} \!+\! \mathbf{Z} \!+\! \sum_{k=1}^{K} t^{-1} \mathbf{H}_{e,k}\mathbf{T}_{e,k}\mathbf{H}_{e,k}^{H}, \nonumber\\
\end{eqnarray} }
We premultiply \eqref{eq:Y_substract_Z} by $ \mathbf{Q}_{s} $,  
{\small  \begin{eqnarray}
  \mathbf{Q}_{s}\bigg(\mathbf{I} \!+\! \mathbf{Z} \!+\! \sum_{k=1}^{K} t^{-1} \mathbf{H}_{e,k}\mathbf{T}_{e,k}\mathbf{H}_{e,k}^{H}\bigg) \!=\! [1 \!+\! f(t)]\mathbf{Q}_{s}\mathbf{H}_{s}\mathbf{T}_{s}\mathbf{H}_{s}^{H}, \!\!\!\!\!\!\!\!\!\!\!\!\!\!\!\!\nonumber\\
\end{eqnarray} }

The following rank relation holds:
{\small  \begin{eqnarray}
\textrm{rank}(\mathbf{Q}_{s}) &\!\!\!\!=&\!\!\!\! \textrm{rank}\bigg[\mathbf{Q}_{s}\bigg(\mathbf{I} \!+\! \mathbf{Z} \!+\! \sum_{k=1}^{K} t^{-1} \mathbf{H}_{e,k}\mathbf{T}_{e,k}\mathbf{H}_{e,k}^{H}\bigg)\bigg] \nonumber\\
 &\!\!\!\!=&\!\!\!\! \textrm{rank} (\mathbf{Q}_{s}\mathbf{H}_{s}\mathbf{T}_{s}\mathbf{H}_{s}^{H} ) \nonumber\\
 &\!\!\!\! \leq &\!\!\!\! \min \{ \textrm{rank}(\mathbf{H}_{s}\mathbf{T}_{s}\mathbf{H}_{s}^{H}), \textrm{rank}(\mathbf{Q}_{s}) \}
\end{eqnarray} }
Based on the above rank relation, we need to show $ \textrm{rank}(\mathbf{H}_{s}\mathbf{T}_{s}\mathbf{H}_{s}^{H}) \leq 1 $ if we claim $ \textrm{rank}(\mathbf{Q}_{s}) \leq 1 $, thus, we consider the following two facts:
{\small \begin{eqnarray}
&& \left[\!\!\begin{array}{cc}
 \mathbf{I}_{N_{T}} \!&\! \mathbf{0}
\end{array}
\!\!\right] \mathbf{H}_{s}^{H} \!=\! \mathbf{I}_{N_{T}}, \nonumber\\
&& \left[\!\!\begin{array}{cc}
 \mathbf{I}_{N_{T}} \!&\! \mathbf{0}  
\end{array}
\!\!\right] \mathbf{A}_{1} \!=\! \beta_{s}\bigg( \mathbf{H}_{s} \!-\! \left[\!\!\begin{array}{cc}
 \mathbf{0}_{N_{T}} \!&\! \mathbf{\bar{h}}_{s}
\end{array}
\!\!\right] \bigg). \nonumber
\end{eqnarray} }
Premultiplying $ \left[\!\!\begin{array}{cc}
 \mathbf{I}_{N_{T}} \!&\! \mathbf{0}
\end{array}
\!\!\right] $ and postmultiplying $ \mathbf{H}_{s}^{H} $ by \eqref{eq:KKT_Main_channel_LMI}, respectively, we have 
{\small  \begin{eqnarray}
&&\!\!\!\!\!\!\!\!\!\!\!\!\!\!\!\!\!\! \beta_{s}\bigg(\mathbf{H}_{s} \!-\! \left[\!\!\begin{array}{cc}
 \mathbf{0}_{N_{T}} \!&\! \mathbf{\bar{h}}_{s}
\end{array} 
\!\!\right]\bigg)\mathbf{T}_{s}\mathbf{H}_{s}^{H} \!+\! [\mathbf{Q}_{s} \!-\! f(t)\mathbf{W}]\mathbf{H}_{s}\mathbf{T}_{s}\mathbf{H}_{s}^{H} \!=\! 0, \nonumber\\
&&\!\!\!\!\!\!\!\!\!\!\!\!\!\!\!\!\!\! \Rightarrow\!\! \bigg(\beta_{s}\mathbf{I} \!+\! [\mathbf{Q}_{s} \!-\! f(t)\mathbf{W}]\bigg)\mathbf{H}_{s}\mathbf{T}_{s}\mathbf{H}_{s}^{H} \!=\! \beta_{s}\left[\!\!\begin{array}{cc}
 \mathbf{0}_{N_{T}} \!&\! \mathbf{\bar{h}}_{s}
\end{array}
\!\!\right]\mathbf{T}_{s}\mathbf{H}_{s}^{H}. \!\!\!\!\!\!\!\!\!\!\!\!\!\! 
\end{eqnarray} } 
\begin{lemma}
If a block hermitian matrix $ \mathbf{P} \!=\! \left[\!\!\begin{array}{cc}
 \mathbf{P}_{1} \!&\! \mathbf{P}_{2}   \\
 \mathbf{P}_{3} \!&\! \mathbf{P}_{4}
\end{array}
\!\!\right] \!\succeq\! \mathbf{0} $, 
then the main diagonal matrices $ \mathbf{P}_{1} $ and $ \mathbf{P}_{4} $ are always PSD matrices \cite{Johnson_B85}. 
\end{lemma}
Now, we can claim $ \beta_{s}\mathbf{I} \!+\! [\mathbf{Q}_{s} \!-\! f(t)\mathbf{W}] \!\succeq\! \mathbf{0} $ and is nonsingular, thus pre(post)multiplying by a nonsingular matrix will not change the matrix rank.
Thus, the following rank relation holds:
{\small \begin{eqnarray}
\textrm{rank}(\mathbf{H}_{s}\mathbf{T}_{s}\mathbf{H}_{s}^{H}) &\!\!\!\!=&\!\!\!\! \textrm{rank}\bigg(\left[\!\!\begin{array}{cc}
 \mathbf{0}_{N_{T}} \!\!&\!\! \mathbf{\bar{h}}_{s}
\end{array}
\!\!\right]\mathbf{T}_{s}\mathbf{H}_{s}^{H}\bigg) \nonumber\\
&\!\!\!\! \leq &\!\!\!\! \textrm{rank}\bigg( \left[\!\!\begin{array}{cc}
 \mathbf{0}_{N_{T}} \!\!&\!\! \mathbf{\bar{h}}_{s}
\end{array}
\!\!\right] \bigg) \!\leq\! 1.
\end{eqnarray} }
\end{appendix}
This completes Theorem \ref{lemma:rank-one_of_robust_sec_rate_max}.~~~~~~~~~~~~~~~~~~~~~~~~~~~~~~~~~~~$ \blacksquare $
\balance
\bibliographystyle{ieeetr}
\bibliography{my_references}

\end{document}